\documentclass[12pt,preprint]{aastex}

\def\ga{\mathrel{\mathpalette\fun >}}

\def\fun#1#2{\lower0.837ex\vbox{\baselineskip0ex\lineskip0.209ex
  \ialign{$\mathsurround=0ex#1\hfil##\hfil$\crcr#2\crcr\sim\crcr}}}

\def\msun{M_\odot}

\def\sles{\lower2pt\hbox{$\buildrel {\scriptstyle <}
   \over {\scriptstyle\sim}$}}
 
\def\sgreat{\lower2pt\hbox{$\buildrel {\scriptstyle >}
   \over {\scriptstyle\sim}$}}

\def\ga{\mathrel{\mathpalette\fun >}}

\begin{document}

 \title{Glimm's Method for Relativistic
%Hydrodynamics%\symbolfootnote[2]{footnote}
 Hydrodynamics\footnote{to
     appear in The Astrophysical Journal, June 1, 2008, v. 679}}

%centerline{\bf to appear in The Astrophysical Journal, June 1, 2008, v. 679}

\shortauthors{CANNIZZO, \ GEHRELS, \& VISHNIAC}
\author{
J.~K.~Cannizzo\altaffilmark{1,2},
N.~Gehrels\altaffilmark{1},
E.~T.~Vishniac\altaffilmark{3}
}
\altaffiltext{1}{NASA/Goddard Space Flight Center, Astrophysics Science Division, Greenbelt, MD 20771}
\altaffiltext{2}{CRESST/Joint Center for Astrophysics, University of Maryland, Baltimore County,
                 Baltimore, MD 21250}
\altaffiltext{3}{Department of Physics and Astronomy, McMaster University,
                 Hamilton, ON L8S 4M1}

\begin{abstract}
We present the results of standard 
one-dimensional test problems
in relativistic hydrodynamics using Glimm's 
(random choice) method, and
compare them to results obtained 
using finite differencing methods.
For problems containing profiles with sharp edges, such 
as shocks, we find Glimm's method    yields global errors
     $\sim1-3$ orders of magnitude smaller than the traditional 
techniques. 
 The strongest differences are seen for problems in which
a shear field is superposed.
For smooth flows,  Glimm's method is inferior to standard methods.
The location of specific features
can be off by up to two grid points
with respect to an exact solution in Glimm's method,  
and furthermore curved states
are not modeled optimally since the method idealizes
solutions as being composed of piecewise constant states.
  Thus 
  although Glimm's method is superior at correctly resolving
  sharp features, especially in the presence of shear,
  for realistic applications
  in which one typically finds smooth flows plus strong gradients or
discontinuities, standard FD methods yield smaller global errors.
  Glimm's method may prove  useful in certain applications
 such as GRB afterglow shock propagation into a uniform medium.
%The fact is that for "real"
%applications (e.g., the propagation of an afterglow shock along an!
%  stratified -i.e., non-uniform- external medium), one finds typically smooth flows plus strong gradients or
%discontinuities. Thus, for most "real" applications, Glimm's code is not competitive with standard FD codes, but
%still, in some cases one may use it (e.g., the same afterglow propagation as mentioned above but moving into a uniform
%medium).
\end{abstract}

%{\it Subject headings:}

\section{Introduction}

Interest in relativistic hydrodynamics has
heightened in recent years due to the 
explosion in the field of gamma-ray bursts
(GRBs $-$ Costa et al. 1997,
van Paradijs et al. 1997, 
Frail et al. 1997,
MacFadyen \& Woosley 1999,
Aloy et al. 2000,
Frail et al. 2001,
Fox et al. 2005,
Gehrels et al. 2005, 
Bloom et al. 2006,
%Zhang et al. 2006,
O'Brien et al. 2006).
 The current paradigm for GRBs involves the
extraction of energy from a newly formed $\sim10\msun$ black
hole and collimation into a relativistic jet, which 
then propagates along the line of sight to the observer.
The emission is thus strongly beamed and Doppler boosted.
The interaction of the jet with the circumstellar medium
produces afterglow.
For Newtonian 
hydrodynamics the density contrast
across a strong shock is given by $\rho_{\rm shock}/\rho_{\rm background}
 = (\Gamma+1)/(\Gamma-1)$,
where $\Gamma$ is the polytropic index;
in relativistic hydrodynamics
 $\rho_{\rm shock}/\rho_{\rm background}
 = (\gamma\Gamma+1)/(\Gamma-1)$,
where $\gamma$ is the Lorentz factor
(Blandford \& McKee 1976).
For putative values $\gamma\simeq10^2-10^3$ thought to be 
required for GRB jets, 
  a relativistic shock can have extremely high density and
be very narrow due to Lorentz contraction.
This poses a severe test for standard finite difference (FD) methods, 
and necessitates adaptive mesh refinement
(Zhang  \& MacFadyen 2006,
  Morsony, Lazzati, \& Begelman 2007).
Adaptive refinement techniques also present challenges, as it has yet to be 
demonstrated that increased levels of refinement on a complex,
multidimensional problem, lead to convergent solutions.
   The desired test of showing that a standard
   performance metric integrated over the computational volume
asymptotes to a constant value with increasing level of refinement
has yet to be carried out
(e.g.,  Zhang  \& MacFadyen 2006).
%  On the other hand, 
%even if there existed a method with perfect resolution,
%for ultrarelativistic problems it would
%not be feasible to have a uniform mesh fine enough
%to provide adequate coverage over the entire computational domain.

Traditional methods for calculating hydrodynamical
evolution of a relativistic fluid have relied on 
finite differencing, i.e., 
discretizing
the differential equations
(Norman \& Winkler 1986, Mart\'i \& M\"uller 2003,
%Dolezal \& Wong 1995,
 Del Zanna \& Bucciantini 2002,
Lucas-Serrano et al. 2004).
 Figure \ref{BM_soln} presents an example of smearing inherent in 
standard FD methods.
 It shows the evolution of
  Lorentz factor $\gamma$ in 
a spherical relativistic blast wave 
  calculation initialized with a
Blandford \& McKee (1976) solution, taking $\gamma_0=5$ initially.
 Each panel shows the same initial conditions, with increasing grid
resolution along the $+x-$direction.
 We use a three dimensional
Cartesian 
 grid
 and utilize  the method described in 
  del Zanna \& Bucciantini (2002). 
  Our implementation of their method is 
detailed in Cannizzo, Gehrels, \& Vishniac (2004).
  Within each panel the number of grid points along the direction of propagation
is increased by a factor of 4.
   In  the fourth panel, for
which there are 64 grid points per small tick mark,
 one can see the clear development of a forward/reverse
shock feature. The inherent smearing behavior of the technique is evident
by comparing successive panels.

\section{Background}

Glimm (1965) presented the theoretical basis for the
random choice, or Glimm's  method.
It relies on first idealizing the solution 
in $(P, \rho, v)$ over $N$ grid points
as consisting of $N$ piecewise constant states, and then solving the 
local Riemann problem $N-1$ times between adjacent grid points.
Second, a random location is selected within a cell, the exact
solution evaluated at that point, and then that is used as the
starting solution for the next time step.
%Early work  on the random-choice method
%was plagued by inadequacies
% in the random number generators utilized.
Chorin (1976, 1977) developed Glimm's method into a numerical
algorithm for problems that could be formulated 
in terms of nonlinear hyperbolic conservation laws.
Sod (1978) reviewed several techniques 
for Newtonian hydrodynamics and found Glimm's method
to be superior in terms of preserving the sharpness of shock edges.
 In the early studies using Glimm's method one sees clear deficiencies
in the solutions, however, both in terms of shock front localization
and overall stability.

 A breakthrough came from Colella (1982) who proposed 
using the van der Corput sequence instead of a standard random
number generator for determining 
 the solution evaluation location with cells in each 
time step. This sequence  is generated by a simple
manipulation of the digits in the binary representation
of consecutive integers.

The application of  Glimm's method to relativistic 
hydrodynamics became possible when
Balsara (1994) and 
Mart\'i \& M\"uller (1996) generalized the solution of
the Riemann solution for relativistic hydrodynamics.
Wen, Panaitescu, \& Laguna (1997)
  used the results of Mart\'i \& M\"uller 
to
implement a  relativistic hydrodynamics
  Glimm's method.
Their study and 
%a follow-up by 
Panaitescu et al. (1997)
are  the only works to date
that employ  Glimm's method
for  relativistic hydrodynamics.

\section{Methodology}

 The basic method is demonstrated in Wen et al. (1997, see their Fig. 2).
  In one half time step the exact solution to the local Riemann
problem is calculated between two grid points, at a position 
determined by the van der Corput sequence.
As explained in Colella (1982),
the sequence is determined by taking the binary 
representation of the positive integers,
$1=1_2$, 
$2=10_2$,
$3=11_2$,
$4=100_2$,
$5=101_2$,
$6=110_2$,
$7=111_2$,
$8=1000_2$, etc.,
and then %transforming the binary digits in each successive
%number by 
flipping the binary digits with respect to the (binary) decimal point,
yielding the sequence 
$a_1 = .1_2 = 0.5$,
$a_2 = .01_2 = 0.25$,
$a_3 = .11_2 = 0.75$,
$a_4 = .001_2 = 0.125$,
$a_5 = .101_2 = 0.625$,
$a_6 = .011_2 = 0.375$,
$a_7 = .111_2 = 0.875$,
$a_8 = .0001_2 = 0.0625$,
etc.
   Note that the sequence alternates between
the two half-unit intervals (0, 0.5) and (0.5, 1),
which helps minimize spurious shock propagation.
Furthermore the series  can be shown to be optimal in terms
of uniform coverage of the unit interval (0, 1).
 The $a_i$ value adopted in a given time step
is the same for all inter-grid points. The exact
solution at a given grid point is alternately taken to be
either the left or right solution between two adjacent grid points. 
The solution is evaluated
in  each alternating half-time step
at a time $(1/2)\Delta t$, where
$\Delta t = n_{\rm CFL}/\Delta x$ with CFL (Courant, Friedrichs, \& Lewy 1967)
number $n_{\rm CFL}=0.5$. 
%  is divided by two for each of these alternating half-time
%steps.
  Thus the pure Glimm's Method effectively adopts a CFL number of 0.5 for the full time step.
%Wen et al. (1997) present a formalism for a hybrid code in which
%one reverts to solving the
%usual hydrodynamical equations if the dynamic range
%in any physical variable between two successive grid points is smaller
%than a certain amount. In this study we use only a pure Glimm formalism.
Although most of our results use a simple one dimensional
Cartesian grid,
Wen et al. (1997) also present geometrical correction terms
for carrying out one dimensional calculations in cylindrical or spherical symmetry.

An important advance since Wen et al. (1997)
are the studies generalizing the relativistic Riemann solution to 
include tangential flow  
(Pons, Mart\'i, \& M\"uller 2000, Rezzolla, Zanotti, \& Pons 2003).
This allows one to extend Glimm's method to problems involving
shear, and to begin to envision a two dimensional Glimm's method.
Pons et al. obtain a solution by solving (1) the jump conditions
across shocks, and (2) a differential equation that comes from a
self-similarity condition along rarefaction waves.
Rezzolla et al. present an integral solution to the equation derived
by Pons et al. and they propose 
  an efficient Gaussian quadrature technique 
for solving it.
 To solve the local Riemann problem between adjacent grid points we use
the publicly available code {\tt RIEMANN\_VT.F} written by
 J.-M. Mart\'i and
 E. M\"uller (cf. Mart\'i \& M\"uller 2003)
which uses the formalism described in Pons et al. (2000)
and Rezzolla et al. (2003).

\section{Testing}

Shock tube problems used in
testing hydrodynamical
codes are a subset of the Riemann problems, for which $v=0$ for all $x$.
One dimensional Riemann problems
 are typically run on a grid such that $0 \le x \le 1$,
and the thermodynamic variables $P$, $\rho$, and $v$ are discontinuous
across $x=0.5$ initially. % ($t=0$).
Starting the simulation is equivalent to removing  a  diaphragm
between left (L) and right (R) states. The strong gradients
across  $x=0.5$ result in four constant states separated by three
elementary waves: rarefaction, contact discontinuity, and shock wave.
Analytical solutions for the time evolution 
 of these problems
for (special) relativistic hydrodynamics are given by Mart\'i \& M\"uller (1994)
for nonshearing problems, 
   and by Pons et al. (2000)
for Riemann problems with added shear (i.e., non-zero $v_\perp$).

 The level of agreement between the exact, analytical solutions and
the numerical ones is quantified by the $L_1$ norm error,
defined for 1D problems as $L_1 = \Sigma_j\Delta x_j |u_j - u(x_j)|$,
where $x_j$ is the coordinate of grid point $j$, $u(x_j)$ is the analytical
value, and $u_j$ the numerical value. The grid spacing is $\Delta x_j$.
 For consistency with previous groups,
we take the solution in proper density. 
The analytical and numerical solutions are calculated on the same grids, and 
the number of grid points $N$ in the solutions are varied
between trials.

\subsection{Riemann Problem 1}

The values in the
initial left and right states are $(p, \ \rho, \ v)_L = (40/3, \ 10, \ 0)$ and
                                  $(p, \ \rho, \ v)_R = ( (2/3)\times 10^{-6}, \ 1, \ 0)$.
 The adiabatic index $\Gamma=5/3$. The result at $t=0.4$
is compared to the analytical one.
The gradient in pressure $p$ produces in the subsequent evolution
a rarefaction wave moving left and a shock wave moving right, with a contact
discontinuity between. The flow is mildly relativistic, with  post-shock velocity  $v=0.714$.
Figure \ref{f.ri1} shows a comparison of the Glimm solution
with the exact one, computed on a grid with $N=400$.
 The small inset panels show a detail of the leading and trailing
edges of the density spike associated with the shock.
For the time step shown, the leading edge of the Glimm solution
is off the analytical solution by one grid point, and the
trailing edge is exact.
Table \ref{t.R1}  presents the $L_1$ errors in density 
between 3 methods, FLASH (from Morsony et al. 2007),
  WENO (weighted essentially nonoscillatory, from
 Zhang \& MacFadyen 2006), and Glimm.
%for Riemann problem 1. 
%The errors for the FLASH code and WENO method (Zhang \& MacFadyen 2006)
%  are taken from  Morsony et al. (2007).
%%
(This test problem has been studied
    by many workers previously 
$-$ e.g., 
    Hawley, Smarr, \& Wilson 1984,
    Schneider et al. 1993,
    Mart\'i \&  M\"uller 1996,
    Wen, Panaitescu, \& Laguna 1997,
    Mart\'i et al. 1997,
    Aloy et al. 1999.)
The asterisked values indicate those trials for which 
%$s=0$.
 the leading and
 trailing shock edge positions
 of the Glimm solutions
    are in agreement with the analytical ones.

For a small sample of individual Glimm trials,
the $L_1$ error is not always a consistent indicator of
success. Although shock propagation speeds
 are expected to be accurate in an averaged sense,   within
a given time step specific features
in   the Glimm solution can be one or two grid points off
from  their
correct location. 
  For problems with sharp edges, such as shocks, the error
will be large (locally) at such a position. Most the rest of the error
is introduced by idealizing the curved state (Riemann fan) to be 
composed of a series of piecewise constant states.
Even if a shock edge location is incorrect at a given time
step,
at a slightly later time step,  or at the same time step for
a run with a different number of grid points $N$, the Glimm solution may
have the correct location of the shock front edges. Therefore a better 
way to measure the success of the method is to plot the $L_1$ errors 
for a large number of different trials, all compared at the same time step 
with the analytical solution for the same $N$. 
%as in the standard tables of comparisons.  
For problems which are typically dominated by one large density enhancement,
one observes bands of solutions
representing those for which the calculated edges are
 (1) exact, (2) off by one grid point
(leading or trailing edge), (3) off by two grid points total, 
(4) off by  three  grid points total, etc.
 We denote the cumulative grid point error 
in shock front localization
%with the analytical solutions
 by $s$.

This effect is shown in Figure \ref{f.R1} where we plot the $L_1$ error
versus $N$. The black and blue points indicate values for the 6 grid points
shown in Table \ref{t.R1}, and the red points show a much larger sample 
drawn from  $\sim10^2$  equi-logarithmically  spaced $N$ values
for the Glimm solutions.
There is a large scatter vertically in the Glimm $L_1$ errors
according to the degree of matching of the shock edges.

\subsection{Riemann Problem 2}

Riemann problem 2
   has a more extreme pressure contrast between the L and R states
initially than problem 1, and therefore drives a faster and higher density
shock.
The values in the
initial left and right states are $(p, \ \rho, \ v)_L = (10^3,  \ 1, \ 0)$ and
                                  $(p, \ \rho, \ v)_R = ( 0.01, \ 1, \ 0)$.  
 The adiabatic index $\Gamma=5/3$. The result at $t=0.4$
is compared to the analytical one.
The flow is relativistic, with post-shock velocity  $v=0.96$.
The shock speed is 0.986.
The width of the shock is $\delta x_s \simeq0.01$ at $t=0.4$ and  for $N=400$ is
covered by 4.2 grid points (in the analytical solution).
 The asterisked values indicate those trials for which $s=0$.
% the leading and
%trailing shock edge positions are in agreement with the analytical ones.
Table 2 compares the $L_1$ errors for the three methods.

Figure \ref{f.R2} shows the  $L_1$ error plot for Riemann problem 2, with 
the values given in Table \ref{t.R2} plus Glimm values for $\sim10^2$ additional $N$ values.
Due to the thinness of the shock compared to problem 1, there is now a clear
banded structure to the Glimm solutions. The lowest striation, which also
contains the first and sixth values from Table \ref{t.R2}, corresponds to solutions for 
which both leading and trailing shock edge positions 
are exact,   $s=0$. The next 
highest striation, containing Glimm entries  $3-5$ from the table, corresponds 
to $s=1$,
and the third striation, containing the second Glimm entry from the table, 
corresponds to $s=2$. The first striation lies about two
orders of magnitude below the F errors, while the second and third are
within a factor $\sim3-10$ of F.

\subsection{Riemann Problem 3}

Riemann problem 3 starts with a strong negative
pressure gradient that launches a reverse shock, and
a positive flow speed in the left state that initiates
a forward shock. Thus there is no Riemann fan.
The values in the
initial left and right states are $(p, \ \rho, \ v)_L = (1,  \ 1, \ 0.9)$ and
                                  $(p, \ \rho, \ v)_R = (10, \ 1, \ 0)$.  
 The adiabatic index $\Gamma=4/3$. The result at $t=0.4$
is compared to the analytical one.
%* The asterisked values indicate those trials for which the leading and
%trailing shock edge positions are in agreement with the analytical ones.
 Table \ref{t.R3} compares the $L_1$ errors for the three methods.

Figure \ref{f.R3}  shows the  $L_1$ error plot for Riemann problem 3, with 
the values given in Table  \ref{t.R3}  plus 
Glimm values for $\sim10^2$ more $N$ values.
None of the values in the table lie in the band 
for $s=1$.
The three upper limit triangles indicate solutions for which
$s=0$, i.e.,  all
three shock edge locations in the problem are exact at $t=0.4$.
  The only limiting precision is the  machine epsilon $\epsilon$ ($\sim 10^{-15}$).
  For one dimensional problems consisting only of constant states,
Glimm's method finds the exact values, therefore the only  error
is introduced by shock edge location inaccuracies.
In traditional methods this test problem produces postshock pressure
oscillations in the reverse shock 
(e.g., Lucas-Serrano et al. 2004, see their Fig. 1; 
         Zhang \& MacFadyen 2006, see their Fig. 3). 
  Lucas-Serrano et al. (2004)
  note, however, that the oscillations completely disappear
when the CFL number  is reduced below 0.3.

\subsection{``Easy'' Shear: Riemann Problem 2 with $(v_\perp)_R \neq 0$}

We now  proceed to one dimensional problems involving shear.
The ``easy'' shear  problem takes
Riemann problem 2 and adds
constant background shear in the R state, $(v_\perp)_R = 0.99$.
 The adiabatic index $\Gamma=5/3$, and the result at $t=0.4$
is compared to the analytical one.
The highest Lorentz factor in the resulting flow $\gamma\sim7.1$.
Unlike purely Newtonian flows in which orthogonal components
of the velocity field are decoupled from each other
(aside from dissipation), with special relativity we now add
the condition that $v^2 + v_\perp^2 < 1$.
   This effectively limits the component of velocity along the
direction of the flow $v$, and also the degree of density enhancement
relative to background within the shock.
  In addition, $\gamma$ now includes a contribution
from the shear.
   There is also a back reaction in terms of the evolution of $v(x,t)$
on the initially  constant $v_\perp$ values.
Table \ref{t.easy} compares the $L_1$ errors for the three methods.

Figure \ref{f.easy}  shows the $L_1$ errors 
for the values given in Table \ref{t.easy},
plus $\sim10^2$ additional $N$ values for G.
As with Figs. \ref{f.R2} and \ref{f.R3},
 the banded structure associated with the
precision in
   the shock edge localization  is evident.
The locus of solutions for $s=0$ 
  lies $\sim10^2 - 10^3$
below the F errors, while the second striation,
 corresponding to
$s=1$,
lies within a factor of 10 of the F errors.

\subsection{``Hard'' Shear: Riemann Problem 2 with $(v_\perp)_R \neq 0$ and  $(v_\perp)_L \neq 0$}

The ``hard'' shear  problem starts with
Riemann problem 2 and adds
background shear in both the R and L states, $(v_\perp)_R = (v_\perp)_L = 0.9$.
 The adiabatic index $\Gamma=5/3$, and the result at $t=0.6$
is compared to the analytical one.
The highest Lorentz factor in the resulting flow is $\gamma\sim35.8$.
Table \ref{t.hard}
compares the $L_1$ errors for the three methods.
 The asterisked value  indicates the trial for which $s=0$.
% the leading and
%trailing shock edge positions are in agreement with the analytical ones.
This problem poses a severe challenge for the traditional methods, but
is well-handled by the Glimm method. In fact, the $L_1$ error for F for
the highest $N$ values shown are equal to those for the lowest $N$ values for G.
Zhang \& MacFadyen (2006) 
 present results of the hard shear test for up to 51,200 grid points,
either uniform or the adaptive mesh equivalent (see their Table 7 and Fig. 9).
 Their $L_1$ errors for $N=51200$ of $\sim 10^{-2}$ are comparable to those 
in our test for $N=400$.
The challenge  of relativistic 1D shearing problems 
for standard FD techniques 
  is also
evident in Morsony et al. (2007, see their Fig. 24).
 The profiles of $\rho$ and $v$ for the FD shearing experiments
shown in Mignone, Plewa, \& Bodo (2005),
         Zhang \& MacFadyen (2006), 
       and Morsony et al. (2007)
all
exhibit a strong displacement and  skewing 
of the shock density spike with respect
 to the analytical solutions.

Figure \ref{f.hard} shows the $L_1$ errors 
for the values given in Table \ref{t.hard},
plus  $\sim10^2$  additional $N$ values for G.
The $s=0$ striation lies $\sim10^2 - 10^3$
below the F errors, and the higher striations are still a factor $\sim10$
below F.

\subsection{Isentropic Smooth Flow}

\subsubsection{Continuous Isentropic}

The previous problems contained sharp gradients produced by shocks.
We now look at a problem with smooth flow, the isentropic flow problem.
This consists of an initial state with smooth profiles in $p$, $\rho$, and $v$.
  A pulse of moving fluid is superposed on top of a constant
density, zero velocity state.
The velocity  of each individual element is constant in time.
Therefore the ``exact'' solution at a later time $t>0$ is
found by advancing each element in time at its known velocity, 
which yields a grid with irregular spacing, and then interpolating
the result back onto a  uniform grid.

%So, how i went about solving this problem is that for any particular
%particle of fluid the velocity doesn't change with time.  So, what i
%did was write a code to set up the initial density and velocity
%profiles on a regularly spaced grid.  I then evolve this grid forward
%in time by adding the velocity of each point by time to find where the
%material at the point ends up.  This gives an irregularly spaced grid,
%but it can be interpolated back to a regularly spaced grid.
%
%So, it's really a semi-analytic solution, but you can make it
%arbitrarily accurate by using a large number of grid points.

The initial structure is given by
\begin{equation}
\rho_0(x) = \rho^{*}[1+\alpha f(x)],
\end{equation}
where $\rho^{*}$ is the density of the constant
background state, and the function
  $f(x)=(x^2 L^{-2}-1)^4$ for  $|x|<L$, and  $f(x)=0$ for  $|x|\ge L$.
 The width of the pulse is $L$ and the amplitude is $\alpha$.
The initial velocity profile within the pulse is
set
  by taking one of the two Riemann invariants to be constant,
 \begin{equation}
 J_{\_} = {1\over 2} \ln \left( {{1+v}\over {1-v}} \right) - {1 \over \sqrt{\Gamma-1} }
   \ln \left( { {\sqrt{\Gamma-1} +c_s} \over {\sqrt{\Gamma-1} -c_s}} \right),
 \end{equation}
where
  $c_s^2 = \Gamma p/(\rho + [\Gamma/(\Gamma-1)]p)$.
The other Riemann invariant is not constant,
 \begin{equation}
 J_{+} = {1\over 2} \ln \left( {{1+v}\over {1-v}} \right) + {1 \over \sqrt{\Gamma-1} }
   \ln \left( { {\sqrt{\Gamma-1} +c_s} \over {\sqrt{\Gamma-1} -c_s}} \right).
 \end{equation}
One inverts the equation for $J_{\_}$ to find the velocity
 \begin{equation}
 v = { {e^{2g}-1}\over{e^{2g}+1}},
 \end{equation}
where
 \begin{equation}
g={J_{\_} + {1 \over \sqrt{\Gamma-1} }\ln\left({\sqrt{\Gamma-1} + c_s} \over{\sqrt{\Gamma-1}-c_s}\right)}.
 \end{equation}
Following previous workers
(Zhang  \& MacFadyen 2006, Morsony et al. 2007)
   we use a domain $-0.35 \le x \le 1$,
and adopt $p^{*}=100$, $\rho^{*}=1$, and $v^{*}=0$.
We also take $\alpha=1$ and $L=0.35$.
 The adiabatic index $\Gamma=5/3$, and the result at $t=0.8$
is compared to the analytical one.
Figure \ref{f.referee5} shows the evolution
of $\rho$, $p$, and $v$ from the initial state.
Table \ref{t.isen}
compares the $L_1$ errors for the three methods.
Figure \ref{f.isen} shows the $L_1$ errors 
for the values given in Table \ref{t.isen},
plus $\sim10^2$ additional $N$ values for G.

\subsubsection{Piecewise Isentropic}

%Along the same lines, I request from the authors to provide a test where a smooth flow is set together with an
%initial discontinuity that should travel over non-uniform initial states. With such a test, the authors would
%provide valuable data on which is the true performance of the Glimm's code for 'realistic' problems (where
%discontinuities and smooth flows are found all together). Such a test will show the ability of the Glimm's method
%to deal with discontinuities between non uniform states. My proposal is that the authors set up an initial
%structure that is piecewise isentropic. Between the two isentropic parts, a jump in  pressure and  in velocity
%parallel to the discontinuity is located, which will evolve into a Riemann fan inmediately.  Comparing the results
%at any given resolution with the solution obtained with the best resolved run (which can be taken as a sort of
%"analytic solution"), the authors may check the order of convergence. 

The Riemann problem and isentropic flow 
problem span extremes of two possible initial states, one
with constant states and one with smooth flow.
 A better metric for realistic problems,
where
discontinuities and smooth flows are found together,
would combine these. 
 Therefore we investigate the evolution
of a structure that is initially piecewise isentropic:
between the two isentropic parts
we introduce a discontinuous jump in pressure
and velocity.
Since there is now no analytical solution, we
carry out one ultra-high resolution run as the reference solution.

The one change we make to the isentropic flow problem is
to force a jump in $p$ at $x=0$ such that the excess above
the floor level $p=100$ drops by a factor of two.
The sharp negative gradient in $p$ at $x=0$ drives a strong flow
to the right which is superposed on the natural flow.
Figure \ref{f.referee6} shows the evolution
of $\rho$, $p$, and $v$ from the initial state,
and Figure \ref{f.referee7} shows the associated errors.
The ``exact'' solution is  obtained by computing a Glimm run for 
$N=10^5$, and then interpolating
to the grid spacing of each of the $\sim10^2$ 
trial runs.
Since this is a modification of a standard test,
there are no FD model errors with which to compare.

\subsection{Shear Suite of Problems from Pons  et al (2000)}

In their generalization of the exact special relativistic
Riemann problem to include shear,
Pons et al. (2000) introduce a suite of 9  tests involving shear,
also based on Riemann problem 2.
These have been examined by Mignone, Plewa, \& Bodo (2005)
using the FLASH code (see their Fig. 5).
In Figure \ref{f.suite} 
we present the results of applying Glimm's method
to this test suite. 
%Following convention we compare numerical
%with analytical results for $N=400$.
  As with the non-shearing test problems, constant states 
 are reproduced exactly (i.e., to within machine precision),
 thereby avoiding the problems
with FD methods alluded to earlier.
%    associated with the shock density spike that are
%       evident in Mignone, Plewa, \& Bodo (2005).

\subsection{Ultrarelativistic Shear Problems from Aloy \& Rezzolla (2006)}

Rezzolla, Zanotti, \& Pons (2003) study  the effect of shear
on the standard Riemann problems, and find that the standard
pattern of a contact discontinuity sandwiched between a 
rightward moving forward shock and a leftward moving reverse shock,
abbreviated ${\mathcal {}_{\leftarrow}S C S_{\rightarrow}}$,
can be fundamentally altered by the presence of a strong shearing field.
For sufficiently large shear, the reverse shock can be replaced by
a rarefaction wave, hence the new pattern 
 ${\mathcal {}_{\leftarrow}R C S_{\rightarrow}}$
 arises.
 Aloy \& Rezzolla (2006) explore the astrophysical ramifications
of the Rezzolla et al finding 
as a potential mechanism for
accelerating jets from AGNs, microquasars, and GRBs to
very high Lorentz factors. They
     show that by varying the left hand pressure $p_L$
in a Riemann problem, one can change the nature of the solution.

We present two additional shearing tests that 
delve deeper into the ultrarelativistic
regime than the ``hard'' shear problem presented earlier.
For the first case we take
                             $(p, \ \rho, \ v, \ \gamma)_L = (10^{-3}, \ 10^{-4}, \ 0.99, \ 20)$ and
                             $(p, \ \rho, \ v, \ \gamma)_R = (10^{-6}, \ 10^{-2}, \    0, \  1)$.
 The Lorentz factor $\gamma$ includes both the normal and perpendicular
velocities $\gamma=(1-v^2-v_\perp^2)^{-1/2}$.
The adiabatic index $\Gamma=4/3$,
      corresponding to the ultrarelativistic case. 
For this trial the shock speed $v_s=0.151$.
The result at $t=1.8$
is compared to the analytical one.
According to Aloy \& Rezzolla (see their Fig. 4),  $p_L=10^{-4}$
 should lie below the transition point from 
${\mathcal {}_{\leftarrow}S C S_{\rightarrow}}$
  to 
 ${\mathcal {}_{\leftarrow}R C S_{\rightarrow}}$.
Figure \ref{f.referee1}
 shows a comparison between 
  the Glimm's Method solution and the exact solution
for $N=400$,
and figure  \ref{f.referee2}
shows the L1 norm density errors at  $t=1.8$.
Since this problem is relatively new, there are no published FD results with 
which to compare, but one suspects that the FD errors 
                     would be comparable or worse to those
shown previously in connection with  the ``hard'' shearing problem.

For the second Aloy \& Rezzolla shear problem we 
increase  $p_L$ by eight orders of magnitude to $10^5$. 
All other initial $L$ and $R$ parameters are the same.
This $p_L$ value should shift
     the wave pattern for the Riemann solution well into
the regime
 ${\mathcal {}_{\leftarrow}R C S_{\rightarrow}}$
and yield a flow with maximum $\gamma\approx 10^3$
(Aloy \& Rezzolla 2006 $-$  see their Fig. 4).
For this trial the shock speed $v_s=0.200$.
The result at $t=0.8$
is compared to the analytical one.
Figure \ref{f.referee3} 
 shows a comparison between 
  the Glimm's Method solution and the exact solution
for $N=400$,
and figure  \ref{f.referee4}
shows the L1 norm density errors at  $t=0.8$.
For large $N$ the Glimm solutions acquire a permanent 
offset error in shock edge localization, rather than 
deviating about a mean $s=0$.
As with Fig. \ref{f.referee2} we have only Glimm errors
to present
   because the
test is too new to have undergone published FD testing.

\subsection{Spherical Blast Wave}

The evolution of a relativistic blast wave in spherical symmetry
has been examined by many workers.
Panaitescu et al. (1997) present a detailed study using a hybrid
Glimm/FD code, and taking $\gamma_0=10^2$.
Kobayashi \& Zhang (2007) utilize a spherically symmetric 
relativistic code which uses a second-order Godunov method
with an exact Riemann solver (described in Kobayashi, Piran, \& Sari 1999)
to investigate the evolution of a relativistic blast wave.
Kobayashi \& Zhang investigate a thin-shell case taking $\gamma_0=10^2$,
and a thick-shell case taking $\gamma_0=10^3$.

The final test shown in Wen et al. (1997) is that for
a relativistic blast wave with initial Lorentz factor $\gamma_0=10$.
For comparison in Figure \ref{KZ_soln}  we show  results
for a run with similar starting conditions.
To adapt to spherical geometry we use the 
geometrical correction terms
given in Wen et al. (1997) with $\alpha=2$.
Within a narrow radial range $0.01r_0$ centered at $r_0$ we initialize
using a Blandford-McKee profile 
$\rho_0(r)=10^4 \gamma_0^2 \chi^{-7/4} \gamma^{-1}$,
where $\chi=1+16(1-r/r_0)\gamma_0^2$, $\gamma=\gamma_0 \chi^{-1/2}$,
$\gamma_0=15$, and $r_0=0.4$. We take 
$p_0(r)=0.2\rho_0(r)$.  
 Inside the initial shell $\rho_0=p_0=10^{-4}$;
outside the initial shell $\rho_0=1$ and $p_0=10^{-4}$.

The profiles shown in 
  Kobayashi \& Zhang (2007) do not display obvious oscillations in the
 shocked shell.
  In our case, using a much smaller initial Lorentz factor, 
we see in Figure \ref{KZ_soln}  
  a number of small oscillations, particularly in $\gamma$.
This indicates that the treatment of
 spherical geometry is worse than that of FD conservative methods
 such as the one of Kobayashi \& Zhang.
 In addition, due to the sharpness of the density shell
and the strong mass jumps accompanying grid points entering into
and then leaving the shell, mass is conserved for
 the run shown in Figure \ref{KZ_soln} only to within $\sim$10\%.

%if(     xa[i]<xcd)
%  if(     xa[i]>0.99*xcd)
%//RBW:
%  {g_fac0 = 15.;

%/ Blandford & McKee:
%    xi = fabs(1. - (xa[i])/0.4) * 2.*g_fac0*g_fac0;
%   chi = 1. + 2.*(3.+1.)*xi;
%    g_fac = sqrt(0.5 * (2.*g_fac0*g_fac0)/chi);
%    if(g_fac < 1.0001) g_fac = 1.0001;

%        dena[i]  =  5200.  *2.*g_fac0*g_fac0 /(pow(chi, 1.75)) /g_fac;  // density  
%        vela[i] = (0.994987437); //  -  0.125 + 0.125*(xcd2-xa[i])/(xcd2-xcd1)); 
%        prea[i] = 0.2*dena[i];

%The evolution of a relativistic blast wave in spherical symmetry
%has been examined by many workers.
%Panaitescu et al. (1997) present a detailed study using a hybrid
%Glimm/FD code.
%Kobayashi \& Zhang (2007) utilize a spherically symmetric 
%relativistic code which uses a second-order Godunov method
%with an exact Riemann solver (described in Kobayashi, Piran, \& Sari 1999)
%to investigate the evolution of a thin relativistic blast wave.
%To adapt to spherical geometry we use the geometrical correction terms
%given in Wen et al. (1997) with $\alpha=2$, and
%in Figure \ref{KZ_soln}  we show for comparison our results
%for a run with
%similar starting conditions.

\section{Discussion}

We have presented the results of a series of 
tests done on standard problems in relativistic hydrodynamics
using Glimm's method.
  To compare to  previous works we utilize the $L_1$ norm errors in 
density. For problems involving smooth gradients such as the
isentropic flow problem, Glimm's method fares worse than the standard
finite difference techniques, due to the fact that solutions are typically
off by $\sim1-2$ grid points. In one dimension, however, the constant
states are exact to within machine precision. 
 This is true irrespective of the presence of shear,
thereby  giving  the method an advantage over FD methods.
If there were only
constant states in a solution, and if the leading and trailing shock
edge locations were correct, then the entire solution would also be
correct (to within machine precision). The idealization of piecewise
constant states for the Riemann fan, however, is a source of error, as
is the incorrect position of a shock edge.
      A better visualization of the Glimm errors than a simple table of
$L_1$  errors versus grid point number $N$ is achieved by 
calculating a large number of numerical and analytical values 
for varying $N$, and plotting the results.
    In such a plot one sees several bands of solutions corresponding
to the total number of grid points $s$  by which the shock edge locations are off.
% added (referee):
   For a given problem, the degree to which sharp edges differ from their
correct locations varies both with time within a given trial,  and  with $N$.
Therefore one cannot choose a priori the ``right''
 resolution for any problem such that the errors are minimized;
  one can only see what the errors are for being off the correct solution
by a given $s$ value.

For the specific problems studied in this work, Riemann problem 1
yields similar global errors between Glimm and FD methods for the ensemble of
$\sim10^2$ solutions.  For  
 Riemann problem 2, the Glimm errors are
comparable to FD for solutions for which 
$s\sim 3-4$. The solutions with 
zero localization error $s=0$ (i.e., exact
matching of the shock edges to their correct values)  have $L_1$ errors $\sim10^2$ 
times smaller than the FD methods.
For Riemann problem 3, the  $s=0$
solutions
are limited only by the
machine $\epsilon$ error, solutions
for which $s=1$ lie a factor $\sim10$
below FD, and solutions with $s\sim2-4$ are comparable to FD.
For the easy shear problem, the $s=0$ solutions have errors  $\sim10^3$ 
times smaller than for FD. The errors become comparable for $s\sim3-4$.
For the hard shear problem, the $s=0$ solutions have errors  $\sim10^2-10^3$
times smaller than for FD. The errors do not become comparable for
any $s$. In fact, 
  the Glimm errors for the lowest $N$ values studied
are comparable to those for the highest $N$ values in previous FD investigations.
For smooth isentropic flow, the FD errors are comparable to Glimm for the smallest $N$
values.
For the largest $N$ values,
the  FLASH errors are a factor $\sim10^{2.5}$ smaller
than for Glimm, and for WENO $\sim10^{5.5}$ times smaller than Glimm.
  For the relativistic blast wave test in spherical geometry (1D),  
  the profiles
  are similar to those of a comparable
  run in Wen et al. (1997, see their Fig. 5).

% CPU discussion: referee:

  For the local Riemann problem, the Riemann solver
 {\tt RIEMANN\_VT.F} decomposes each solution into a left
wave and a right wave.
Depending on the conditions, many iterations may be required,
therefore the computation time can varying greatly.
Wen et al. (1997) discuss the slowness inherent 
  in the Glimm's Method  and quote
run times $\ga10$ times slower than standard FD methods.
We find, using a $\sim2$GHz machine that, for example, Riemann problem 1 for $N=400$
and $t=0.4$ (640 half time steps)
requires 7s of CPU time ($27\mu s$ per grid point per half-time step),
the hard shear problem for $N=400$
and $t=0.6$ (960 half time steps)
uses 17s ($44\mu s$ per grid point per half-time step),
 and the Aloy \& Rezzolla problem 2 for $N=400$
and $t=0.8$ (1280 half time steps)
takes 230s ($450\mu s$ per grid point per half-time step).
The $N=10^5$ piecewise isentropic run required 4 wks.

Although Glimm's method is superior in resolving shocks,
for problems containing thin features, as is common in relativistic
hydrodynamics, there is still a strong need for  adaptive mesh refinement.
For a given grid spacing, features are often too narrow to be
resolved.
Figure \ref{mass_ri2} shows the variation of total grid mass $m$ (computed
from the proper density)
with time for Riemann problem 2 for the six Glimm  runs
indicated in Table \ref{t.R2}.
(The ending time $t=0.4$ is that for which the errors were calculated.)
 The abrupt vertical excursions in $m$ arise as the shock widens with
time and new grid points are incorporated into the shock feature.
Since the density  is higher within the shock, the mass jumps.
For the higher $N$ values there are always enough grid points to 
cover the shock, and the variation in total mass is small as the
new shock grid points come into existence.
 For the lower $N$ values, however, this is not the case.
In fact, for the $N=100$ run, there are no grid points representing the
shock feature until $t\simeq0.3$, at which time the shock has widened
to of order the grid spacing, and one grid point appears at the shock
location, hence the large jump in mass.
   For Glimm's method to be a useful research tool, 
it will probably be necessary 
   not only to have a two dimensional version,
but also to include a provision for adaptive mesh refinement.
 Preliminary work on a 2D version has been encouraging, but more
effort is required to address the issue of numerical stability.

\section{Conclusions}

We present the results of relativistic hydrodynamical tests
using Glimm's method, along with a comparison to results
using standard methods.
 Glimm's method in one dimension is superior to standard 
finite differencing for  problems containing shocks,
in  which a sharp gradient appears.
The introduction of shear does not degrade the quality of
the solutions. Indeed, the work of Pons et al. (2000) 
generalizing the relativistic Riemann solution to include shear
now also provides impetus for making a two dimensional 
relativistic Glimm's method.
For problems involving smooth flow, 
the standard finite differencing methods are much better.
Although constant states are 
  calculated  exactly (i.e., to within machine precision)
in Glimm's method, curved states such as Riemann fans
are somewhat imprecisely modeled as being composed of a
sum of piecewise constant states. Furthermore, the 
fact that there is an uncertainty of $1-2$ grid points
in the location of a given feature means that for models
 with smoothly varying physical parameters, the entire
profile can be shifted slightly, leading to large global
errors in comparison to an exact solution.
  The results
  of the piecewise isentropic run
  indicate that
  for realistic applications 
  containing both smooth flows and sharp gradients,
  standard FD methods give superior global behavior.
  Glimm's method may prove better for applications
 such as GRB afterglow shock propagation into a uniform medium
  where one is primarily interested in the physical evolution of
  high entropy
material only within a restricted volume
  (i.e., the shocked gas), and not the global evolution
of low density, low entropy regions far away from the shock.

\acknowledgements

We thank Alin Panaitescu for helpful discussions 
concerning Glimm's method, and for
allowing us to use the driver code from 
Wen, Panaitescu, \& Laguna (1997) that 
  sets up the hydrodynamical model and calls the
Riemann solver. As mentioned earlier, we use the 
code {\tt RIEMANN\_VT.F} written by Jose Mart\'i and Ewald M\"uller 
for solving the Riemann problem.
We also thank Brian Morsony for useful advice and a short IDL code
to advance the exact solution for the isentropic flow problem,
and Tod Strohmayer for a useful suggestion.
  Thanks also go to the anonymous referee for suggesting the
piecewise isentropic flow problem
 and the
   ultrarelativistic shearing problems
from Aloy \& Rezzolla (2006).
%Finally, we thank Jose Font for suggesting this work.

%We acknowledge assistance from the
%{\it Swift} team.

%\epsfbox[70 0 612 662]{tab1.ps}
%\epsfbox[70 0 612 662]{tab2.ps}

%\epsfbox[70 0 612 662]{tables.ps}

\clearpage

\begin{deluxetable}{rlll}
\tablecaption{L1 Error $-$ Riemann Problem 1}
\tablewidth{0pt}
\tablehead{
\colhead{$N$} & \colhead{FLASH} & \colhead{WENO}  & \colhead{Glimm} 
} 
\startdata
100  & 0.13   & 0.13    & 0.029$^*$  \cr
200  & 0.070  & 0.074   & 0.034      \cr
400  & 0.036  & 0.033  & 0.017       \cr
800  & 0.018  & 0.021  & 0.0035$^*$  \cr
1600 & 0.0085  & 0.010  & 0.0033    \cr
3200 & 0.0043  & 0.0051  & 0.0069   \cr
\enddata
\label{t.R1}
\end{deluxetable}

\clearpage

\begin{deluxetable}{rlll}
\tablecaption{L1 Error $-$ Riemann Problem 2}
\tablewidth{0pt}
\tablehead{
\colhead{$N$} & \colhead{FLASH} & \colhead{WENO}  & \colhead{Glimm} 
} 
\startdata
100  & 0.21   & 0.21   & 0.0034$^*$  \cr
200  & 0.15   & 0.14   & 0.10        \cr
400  & 0.083  & 0.093  & 0.024       \cr
800  & 0.046  & 0.055  & 0.012       \cr
1600 & 0.025  & 0.025  & 0.0061      \cr
3200 & 0.013  & 0.015  & 0.00011$^*$ \cr
\enddata
\label{t.R2}
\end{deluxetable}

\clearpage

\begin{deluxetable}{rlll}
\tablecaption{L1 Error $-$ Riemann Problem 3}
\tablewidth{0pt}
\tablehead{
\colhead{$N$} & \colhead{FLASH} & \colhead{WENO}  & \colhead{Glimm} 
} 
\startdata
100  & 0.059  & 0.10   & 0.061  \cr
200  & 0.035  & 0.063  & 0.031  \cr
400  & 0.021  & 0.030  & 0.013  \cr
800  & 0.013  & 0.017  & 0.0070 \cr
1600 & 0.085  & 0.095  & 0.0038 \cr
3200 & 0.033  & 0.052  & 0.0019 \cr
\enddata
\label{t.R3}
\end{deluxetable}

\clearpage

\begin{deluxetable}{rlll}
\tablecaption{L1 Error $-$ Easy Shear}
\tablewidth{0pt}
\tablehead{
\colhead{$N$} & \colhead{FLASH} & \colhead{WENO}  & \colhead{Glimm} 
} 
\startdata
100  & 0.63    & 0.76    & 0.24   \cr
200  & 0.34    & 0.39    & 0.12   \cr
400  & 0.17   & 0.23     & 0.059  \cr
800  & 0.084  & 0.12     & 0.029  \cr
1600 & 0.044   & 0.066   & 0.015  \cr
3200 & 0.023   & 0.034   & 0.029  \cr
\enddata
\label{t.easy}
\end{deluxetable}

\clearpage

\begin{deluxetable}{rlll}
\tablecaption{L1 Error $-$ Hard Shear}
\tablewidth{0pt}
\tablehead{
\colhead{$N$} & \colhead{FLASH} & \colhead{WENO}  & \colhead{Glimm} 
} 
\startdata
100  & 0.51    & $-$     & 0.038   \cr
200  & 0.46    & $-$     & 0.019   \cr
400  & 0.33   & 0.52     & 0.0096  \cr
800  & 0.22   & 0.36     & 0.00048$^*$  \cr
1600 & 0.13    & 0.23    & 0.0030  \cr
3200 & 0.083   & 0.13    & 0.0029  \cr
6400 & 0.053   & 0.065   & 0.0013  \cr
\enddata
\label{t.hard}
\end{deluxetable}

\clearpage

\begin{deluxetable}{rlll}
\tablecaption{L1 Error $-$ Isentropic Flow}
\tablewidth{0pt}
\tablehead{
\colhead{$N$} & \colhead{FLASH} & \colhead{WENO}  & \colhead{Glimm} 
} 
\startdata
  80 & 5.5e-3 & 2.1e-3 & 0.0072  \cr
 160 & 1.6e-3 & 1.1e-4 & 0.0052  \cr
 320 & 4.0e-4 & 1.7e-5 & 0.0033  \cr
 640 & 1.0e-4 & 1.5e-6 & 0.0024  \cr
1280 & 2.5e-5 & 1.6e-7 & 0.0019  \cr
2560 & 5.4e-6 & 1.9e-8 & 0.0014  \cr
5120 & 1.6e-6 & 2.4e-9 & 0.00053 \cr
\enddata
\label{t.isen}
\end{deluxetable}

\clearpage

%%\centerline{\bf Figure Captions}

\figcaption% 1 (new 1)
{The evolution of Lorentz factor $\gamma$ for a Blandford-McKee
initial state with $\gamma_0=5$ in a 3D Cartesian calculation using the
method described in del Zanna \& Bucciantini (2002),
using the local Lax-Friedrichs flux.
The four panels show increasing grid resolution
in a slice along
the propagation direction.
The initial step is the leftmost profile in each panel,
and profiles moving to the right show the
shock development at eight subsequent time steps.
For ease of viewing, the solutions in the first panel are connected
by  solid lines.
The dotted curve in each panel indicates the
$\gamma$ value corresponding to the local maxima in $\rho$ for the nine
time steps. (For the first two panels there is a [spurious]
offset between the local maxima in $\rho$ and $\gamma$.)
The number of grid points per small tick
mark is ({\it top to bottom}) 1, 4, 16,
and 64, respectively.\label{BM_soln}}

\figcaption% 2 (new 2)
{A comparison of the pressure $p$ ({\it open triangles}),
 density $\rho$ ({\it open squares}),
and velocity $v$ ({\it open pentagons}) for Riemann problem 1 at $t=0.4$
using Glimm's method
with the exact solution ({\it solid line}),
 where both are computed on a grid with $N=400$.
The small panels show the detail of the leading and
trailing edges of the density spike.
A dashed line connects the Glimm density points.\label{f.ri1}}

\figcaption% 3 (new 3)
{The $L_1$ errors in density for the F
(FLASH and WENO  $-$ shown in black) and G solutions (Glimm)
in Table \ref{t.R1} $-$ Riemann problem 1 (shown in blue), as well as the results using 
$\sim10^2$ additional $N$ values for G
(shown in red).\label{f.R1}}

\figcaption% 4 (new 4)
{The $L_1$ errors in density for the F (FLASH and WENO  $-$ black)  and G (Glimm $-$ blue) solutions 
in Table \ref{t.R2} (Riemann problem 2), as well as $\sim10^2$ additional $N$ values for G (red).
The fact that the shock is narrower than for Riemann problem 1
 leads to a more pronounced striationing; with fewer points spanning
the shock, the relative error introduced by being off a given number of grid
points in the shock edge location is larger.\label{f.R2}}

\figcaption% 5 (new 5)
{The $L_1$ errors in density for the F (FLASH and WENO  $-$ black)  and G (Glimm $-$ blue) solutions 
in Table \ref{t.R3} (Riemann problem 3), as well as  $\sim10^2$ more $N$ values for G (red).
The solutions listed in the table all lie in the band
for which $s=2$.
The upper limit triangles indicate three solutions which 
are limited only by machine precision ($\sim10^{-15}$).\label{f.R3}}

\figcaption% 6 (new 6)
{The $L_1$ errors in density for the F (FLASH and WENO  $-$ black) and G (Glimm $-$ blue) solutions 
in Table \ref{t.easy} (``easy'' shear), as well as  $\sim10^2$ more $N$ values for G (red).\label{f.easy}}

\figcaption% 7 (new 7)
{The $L_1$ errors in density for the F (FLASH and WENO  $-$ black) and G (Glimm $-$ blue) solutions 
in Table \ref{t.hard} (``hard'' shear), as well as $\sim10^2$ more 
$N$ values for G (red).\label{f.hard}}

\figcaption% 8 (new 8)
{The evolution of $\rho$ ({\it top panel}),
    $p$ ({\it middle panel}),
   and 
    $v$ ({\it bottom panel})
  for the isentropic flow problem,
taking $N=400$.
   Shown are the initial state ({\it dotted}) and 5 subsequent
equally spaced times steps up to $t=0.8$.\label{f.referee5}}

\figcaption% 8 (new 9)
{The $L_1$ errors in density for the F (FLASH $-$ black), W (WENO $-$ black),
     and G (Glimm $-$ blue) solutions 
in Table 6 (isentropic flow), as well as $\sim10^2$ additional
$N$ values for G (red).\label{f.isen}}

\figcaption% 8 (new 10)
{The evolution of $\rho$ ({\it top panel}),
    $p$ ({\it middle panel}),
   and
    $v$ ({\it bottom panel})
  for the piecewise isentropic flow problem,
taking $N=400$.
   Shown are the initial state ({\it dotted}) and 5 subsequent
equally spaced times steps up to $t=0.8$.\label{f.referee6}}

\figcaption% 8 (new 11)
{The $L_1$ errors in density for the 
      Glimm solutions for the piecewise isentropic flow problem.\label{f.referee7}}

\figcaption% 9 (new 12)
{Comparison of numerical with analytical solutions for the 
shearing test problems shown in 
Pons et al.    (2000, see their Fig. 4) and
Mignone et al. (2005, see their Fig. 5)
using $N=400$.
The small insets in each panel show details of 
the leading and trailing edges of the density spike.
The tests begin with Riemann problem 2, and then add,
for the initial shear in the R and L states
({\it from left to right}),
$(v_{\perp})_R=$ 0, 0.9, and 0.99,
and
({\it from top to bottom})
$(v_{\perp})_L=$ 0, 0.9, and 0.99.
The polytropic index $\Gamma=5/3$, and the solution
is evaluated at $t=0.4$.
Thus the upper left panel shows the solution
for
 Riemann problem 2 from Section 4.2, 
  the upper right panel shows 
the ``easy'' shear solution from Section 4.3,
and
the central panel shows
the ``hard'' shear solution from Section 4.4
(evaluated at $t=0.4$, however, rather than $t=0.6$).
For the central and 
lower right panels, $s=0$;
  for all other panels, $s=1$.\label{f.suite}}

\figcaption% 10 new (new 13)
{A comparison of the pressure $p$ ({\it open triangles}),
 density $\rho$ ({\it open squares}),
velocity $v$ ({\it open pentagons}),
and Lorentz factor ($=[1-v^2-v_\perp^2]^{-1/2}$ $-$ {\it open hexagons})
 for the first Aloy \& Rezzolla ultrarelativistic problem at $t=1.8$
%using Glimm's method
with the exact solution ({\it solid line}),
 where both are computed with $N=400$.
The small insert panel shows a detail of
                   the density spike.
A dashed line connects the Glimm density points.\label{f.referee1}}

\figcaption% 11 new (new 14)
{The $L_1$ errors in density accompanying the Glimm solutions
for the first Aloy \& Rezzolla ultrarelativistic problem, at $t=1.8$.
\label{f.referee2}}

\figcaption% 12 new (new 15)
{A comparison of the pressure $p$ ({\it open triangles}),
 density $\rho$ ({\it open squares}),
velocity $v$ ({\it open pentagons}),
and Lorentz factor ($\gamma=[1-v^2-v_\perp^2]^{-1/2}$ $-$ {\it open hexagons})
 for the second Aloy \& Rezzolla ultrarelativistic problem at $t=0.8$
%using Glimm's method
with the exact solution ({\it solid line}),
 where both are computed with $N=400$.
A snapshot of the first three variables is shown in 
the top panel, and the bottom panel shows $\gamma$.
The small insert within the top panel 
 presents a detail of 
           the density spike.
A dashed line connects the Glimm density points.\label{f.referee3}}

\figcaption% 13 new (new 16)
{The $L_1$ errors in density accompanying the Glimm solutions
for the second  Aloy \& Rezzolla ultrarelativistic problem, at $t=0.8$.
\label{f.referee4}}

\figcaption% (old 10) (new 17)
{The evolution of ({\it from top to bottom})
Lorentz factor $\gamma$, 
density $\rho$,
and pressure $p$ for a
spherically symmetric  1D test run
 of a thin-shell, relativistic blast wave
%o compare with Kobayashi \& Zhang (2007, see their Fig. 3).
to compare with Wen et al (1997, see their Fig. 5).
For this run $N=10^5$ over the entire computational domain
($0.075 < r < 5.1$), or 3800 grid points over the domain plotted.
A Blandford-McKee profile with
Lorentz factor $\gamma_0=15$
  is taken initially
for a thin spherical shell extending from $0.99r_s$ to $r_s$,
where $r_s=0.4$. 
The frame of reference is
continually adjusted so that the  origin
corresponds to the position of the contact discontinuity.
There is a rightward moving forward shock and
a leftward moving reverse shock.\label{KZ_soln}}

\figcaption % (old 11) (new 18)
{The variation of total mass with time, integrated over the grid,
for Riemann problem 2.
The six panels accompany the six Glimm entries in Table \ref{t.R2}.\label{mass_ri2}}

\clearpage
\begin{figure}
   \plotone{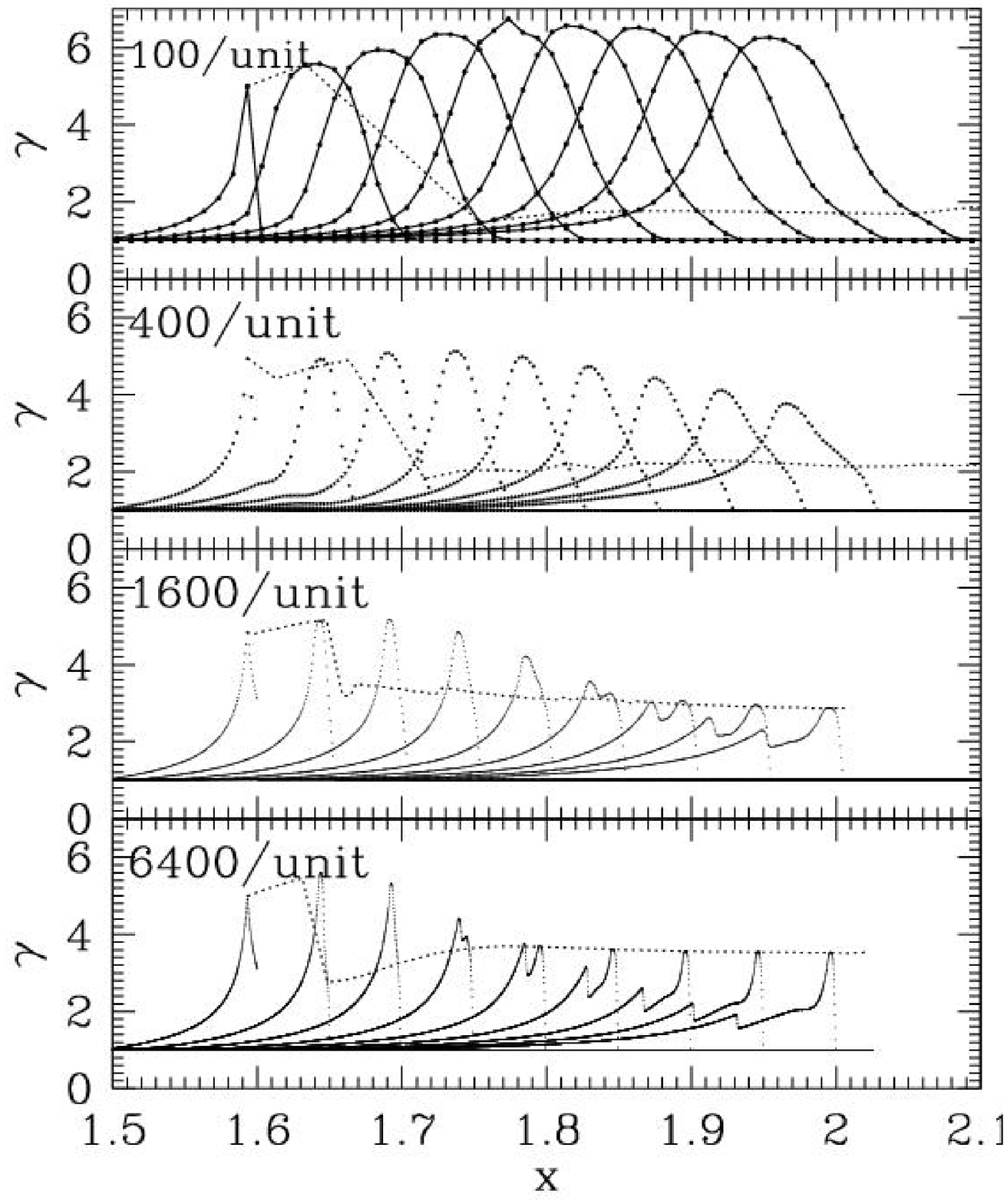}
\centerline{f1.ps}
\end{figure}
\clearpage

\begin{figure}
   \plotone{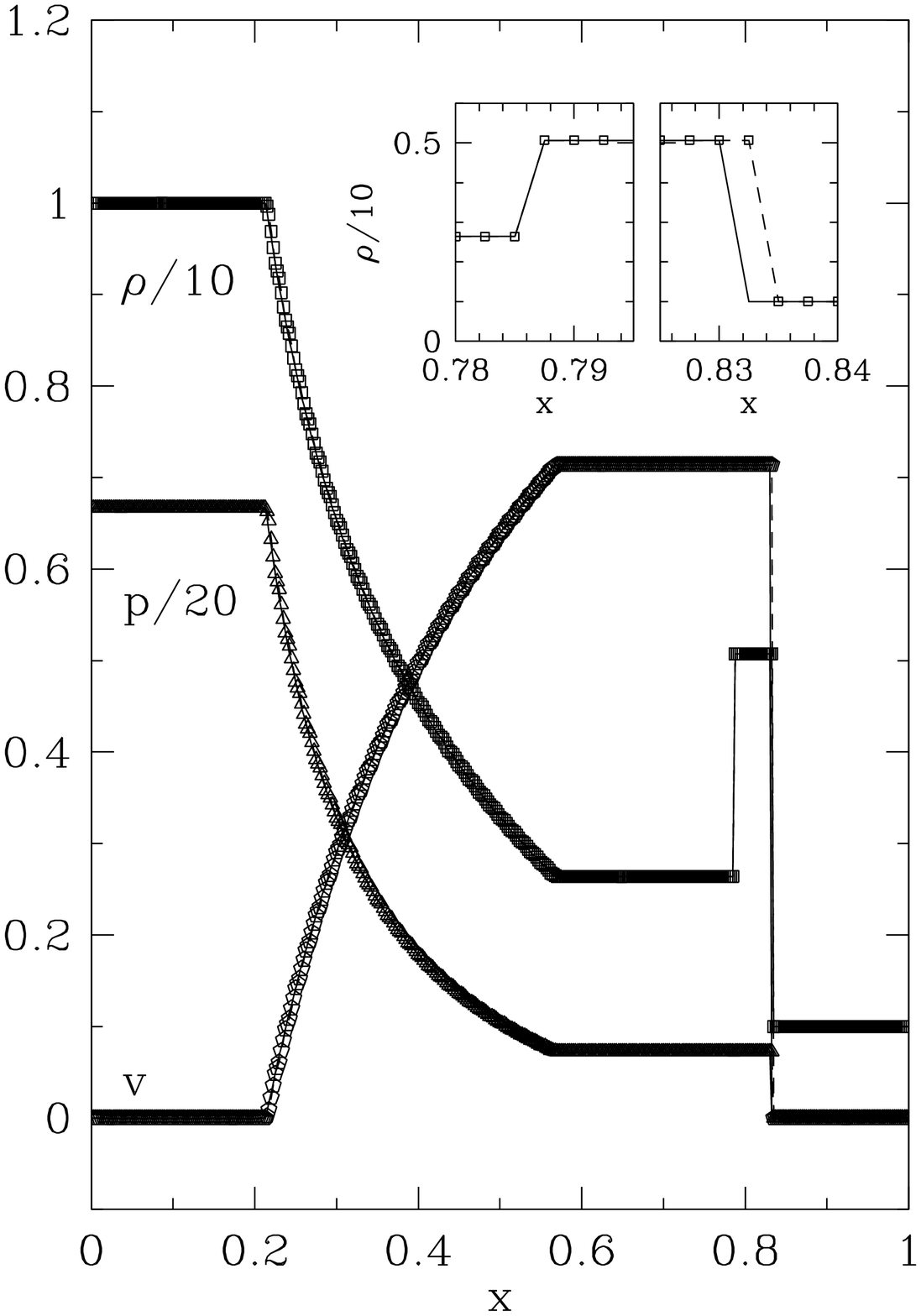}
\centerline{f2.ps}
\end{figure}
\clearpage

\begin{figure}
   \plotone{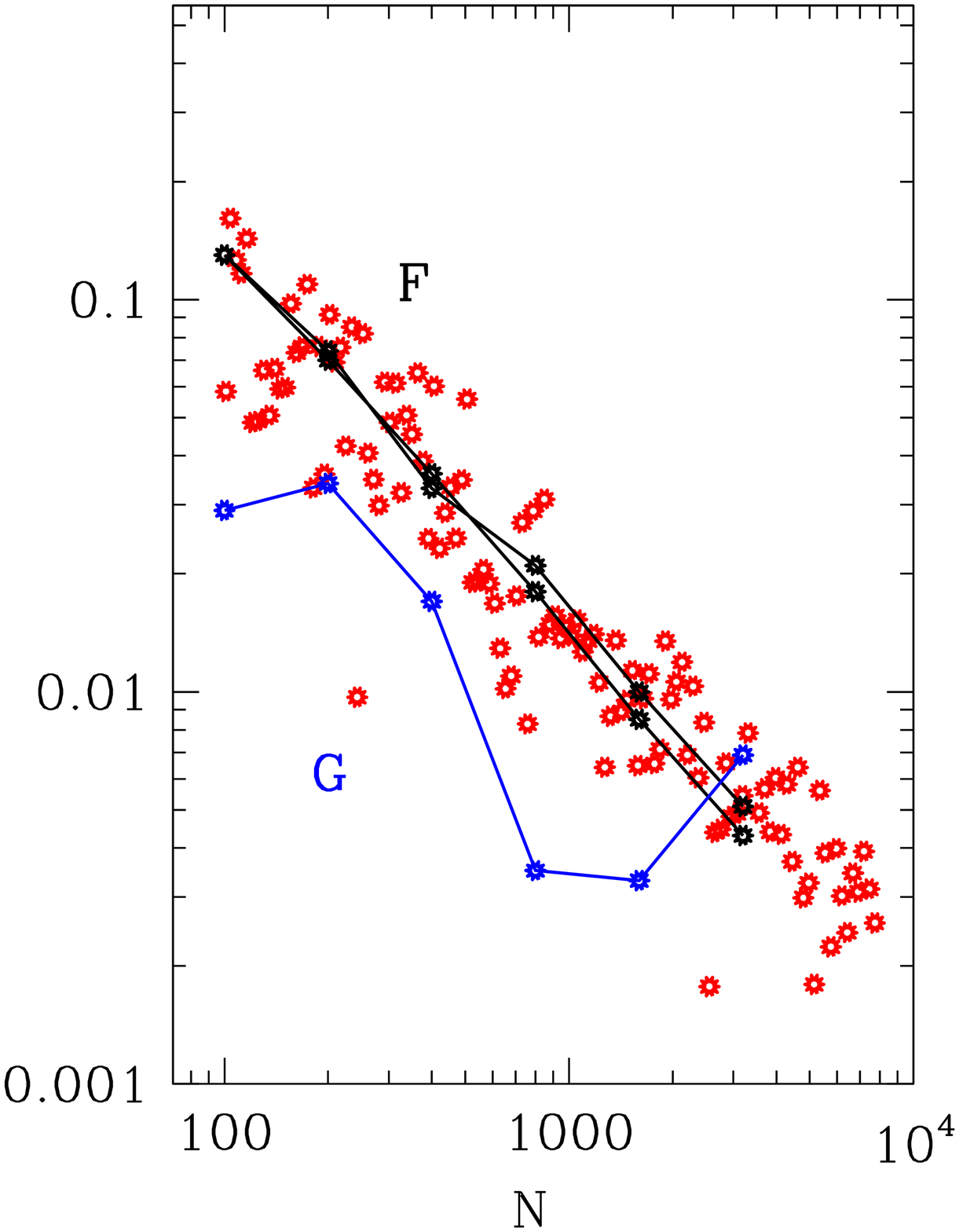}
\centerline{f3.ps}
\end{figure}
\clearpage

\begin{figure}
   \plotone{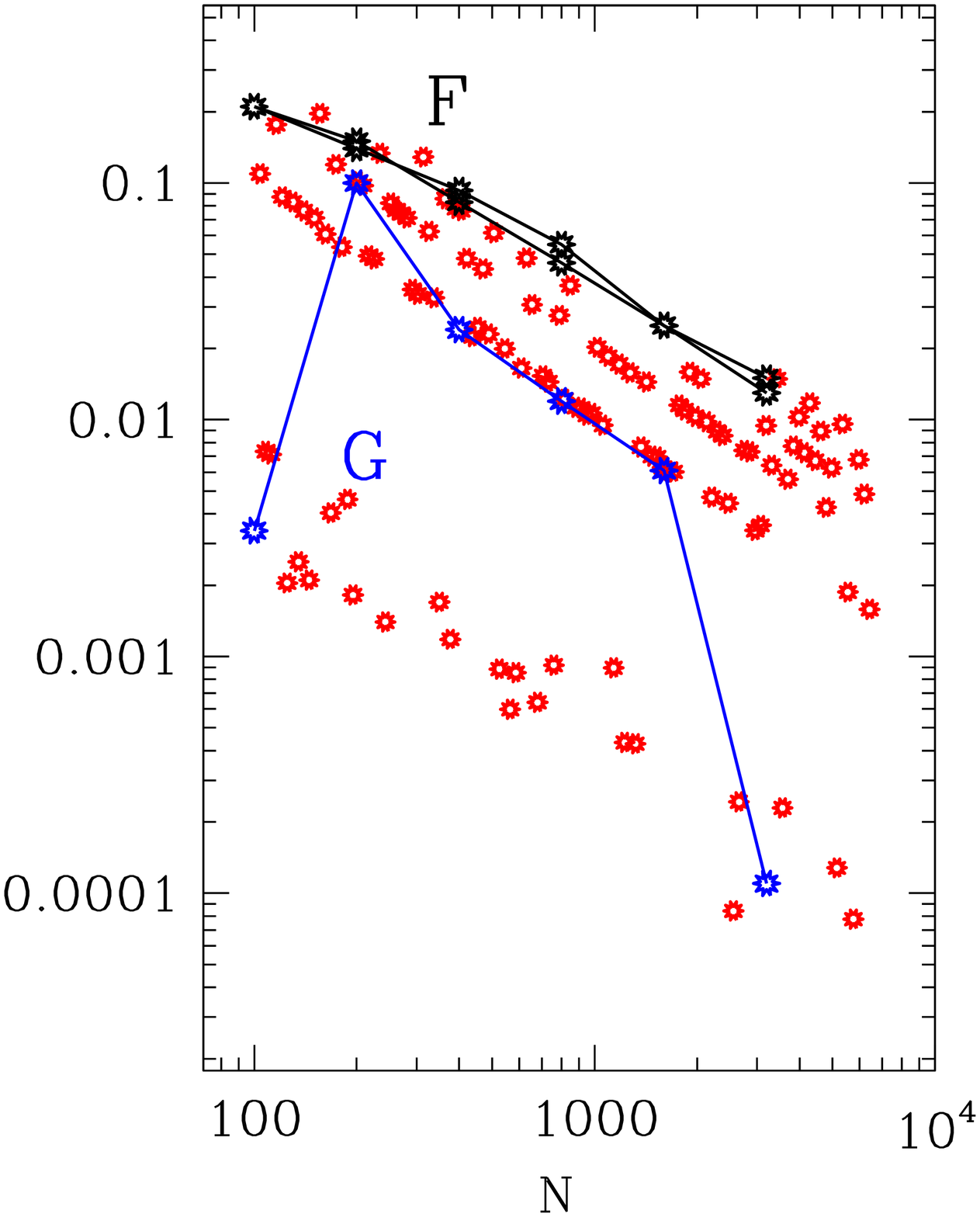}
\centerline{f4.ps}
\end{figure}
\clearpage

\begin{figure}
   \plotone{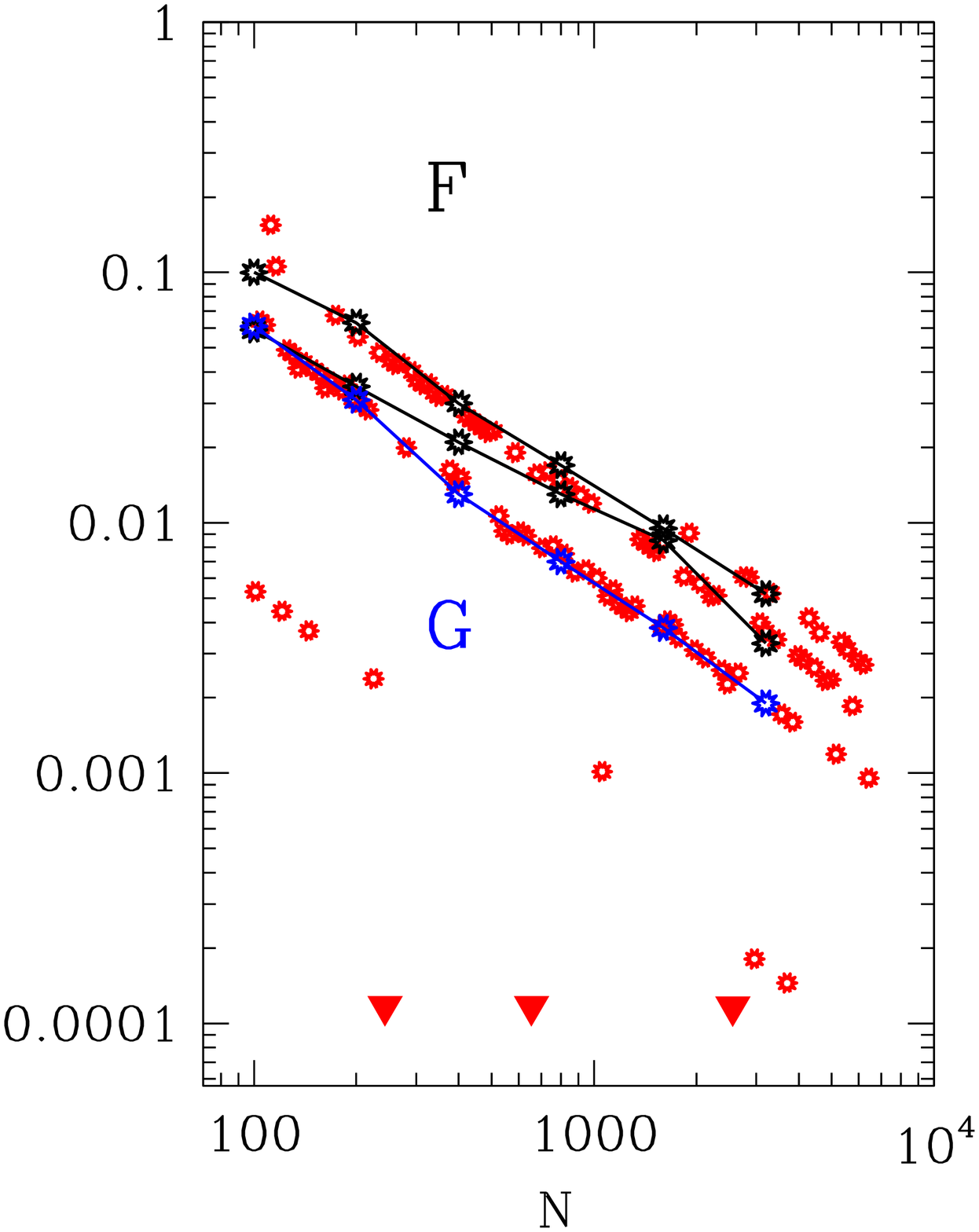}
\centerline{f5.ps}
\end{figure}
\clearpage

\begin{figure}
   \plotone{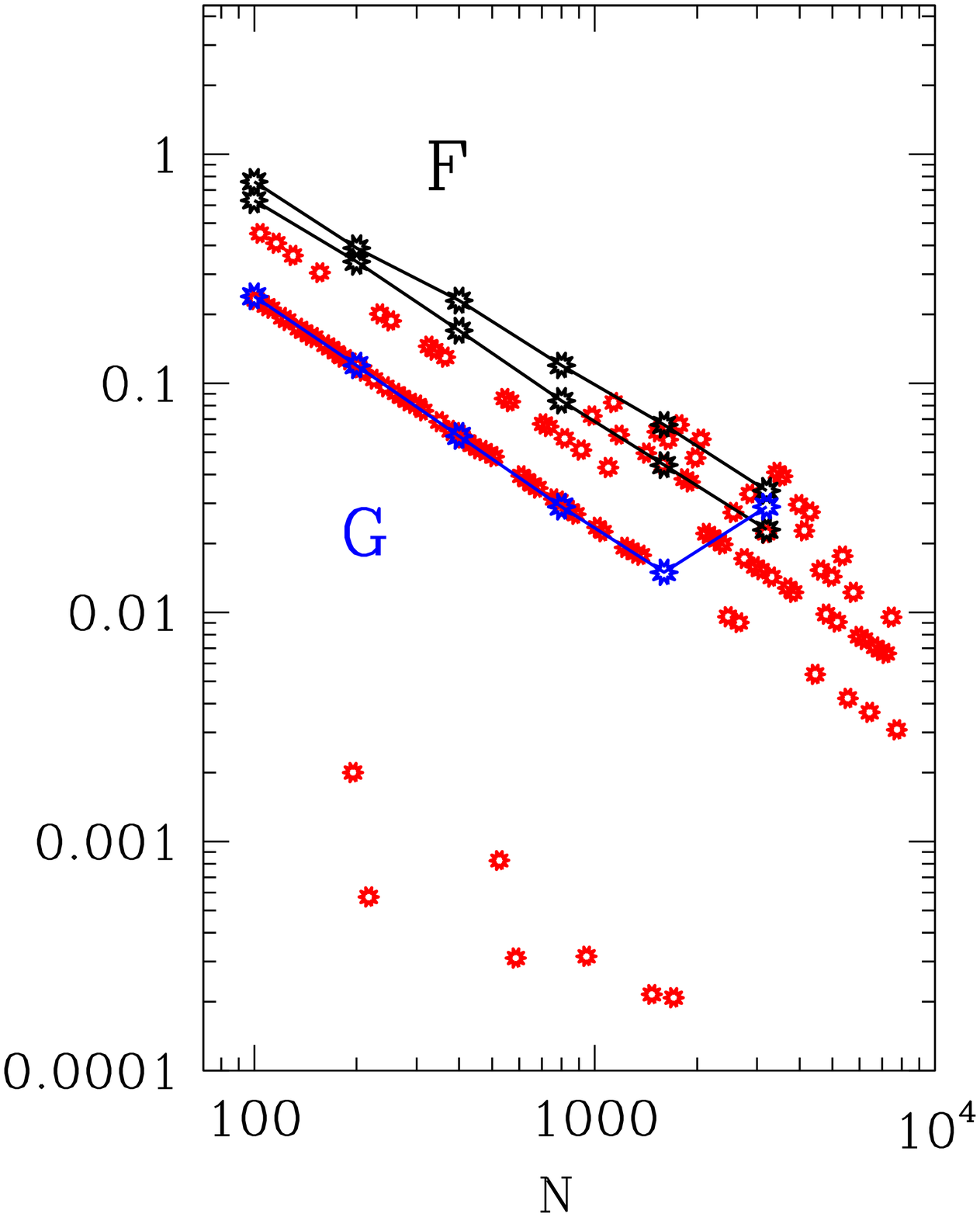}
\centerline{f6.ps}
\end{figure}
\clearpage

\begin{figure}
   \plotone{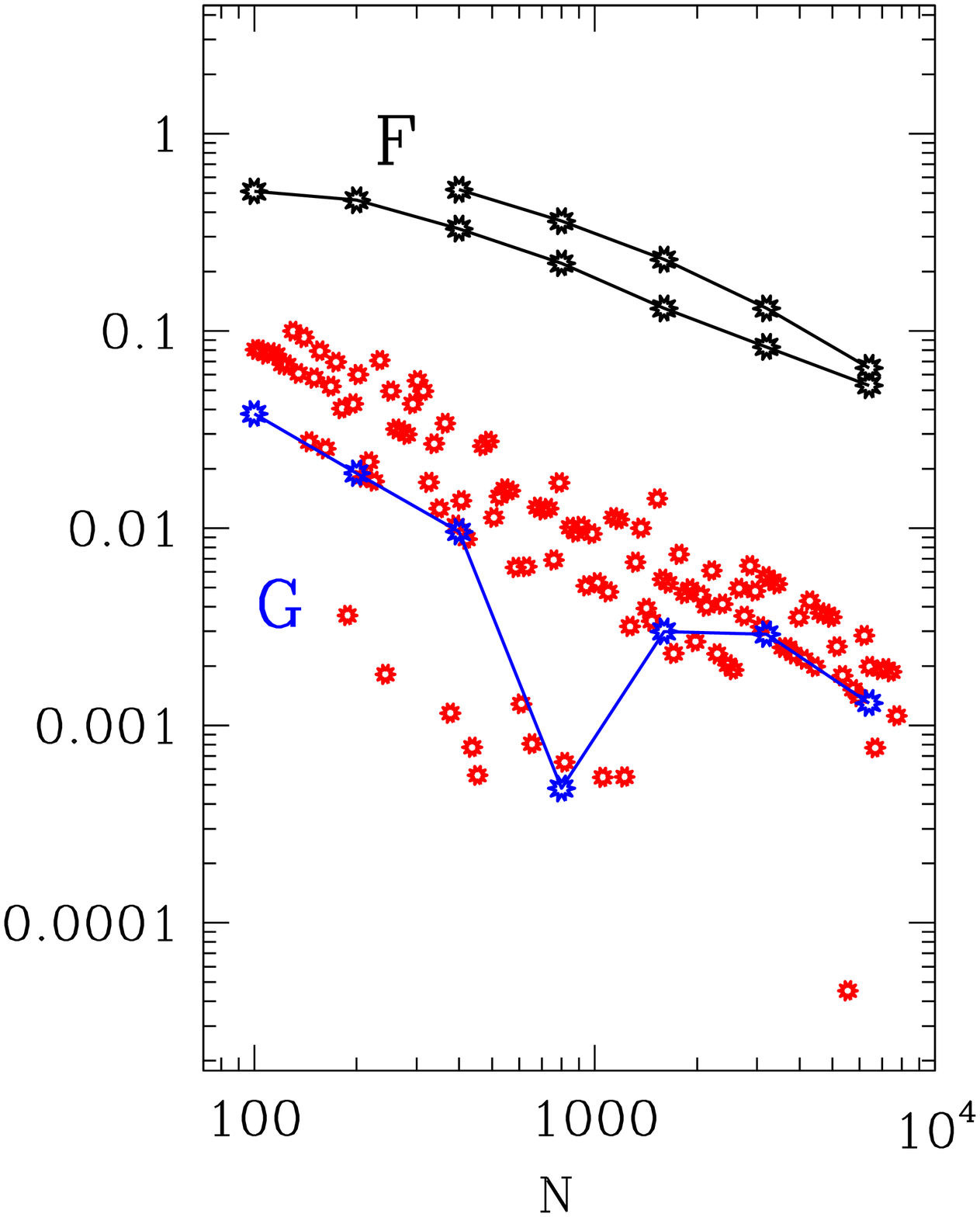}
\centerline{f7.ps}
\end{figure}
\clearpage

\begin{figure}
   \plotone{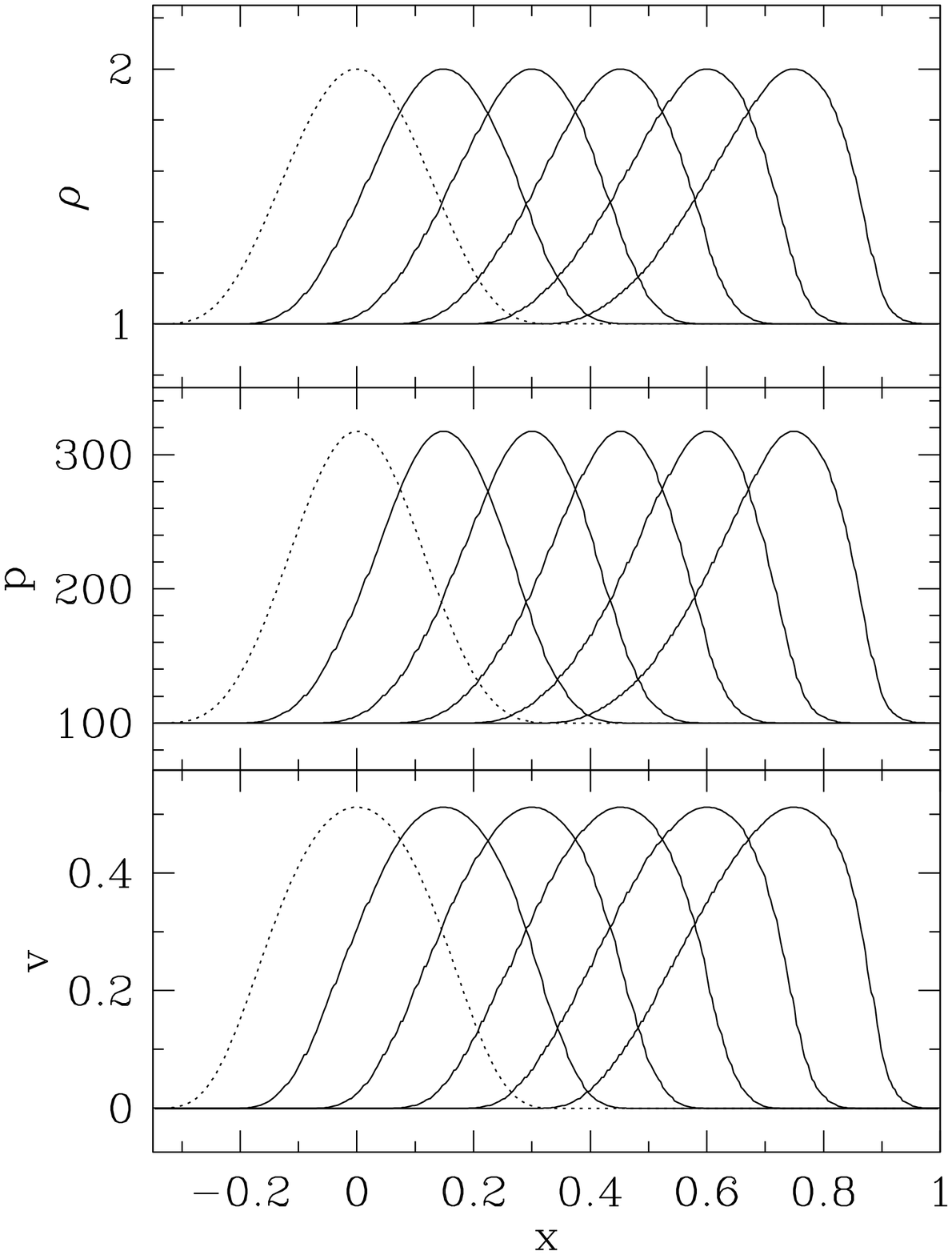}
\centerline{f8.ps}
\end{figure}
\clearpage

\begin{figure}
   \plotone{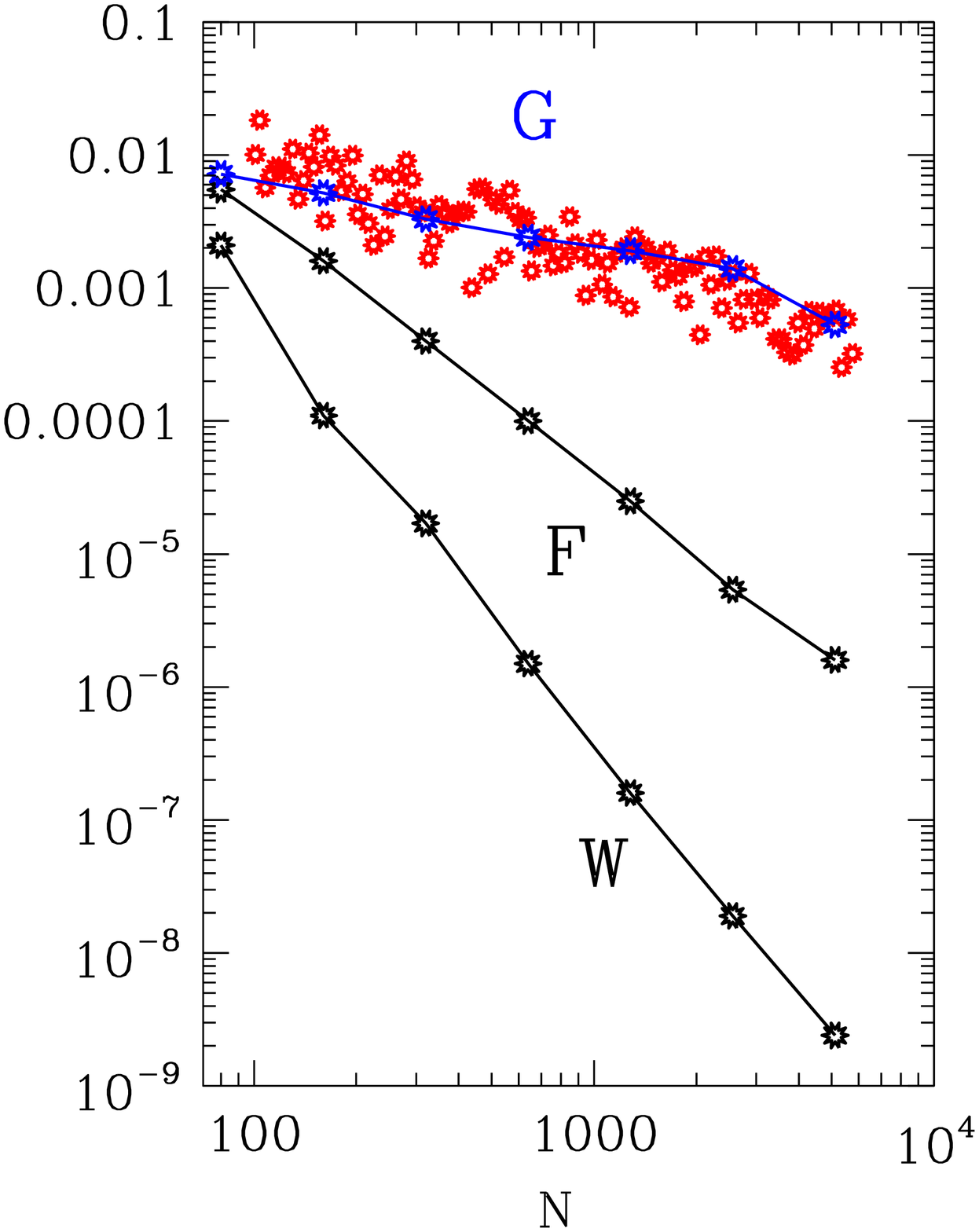}
\centerline{f9.ps}
\end{figure}
\clearpage

\begin{figure}
   \plotone{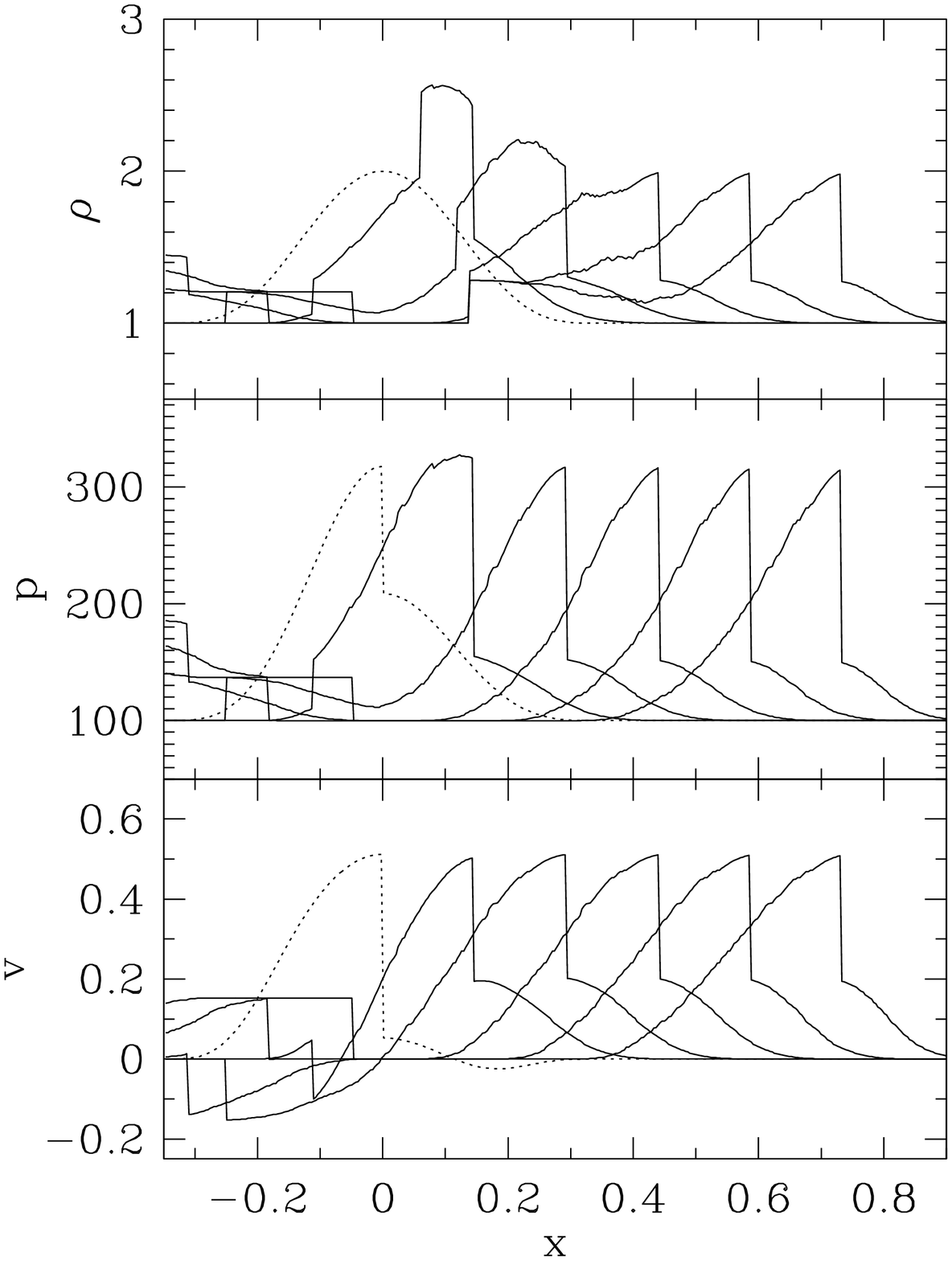}
\centerline{f10.ps}
\end{figure}
\clearpage

\begin{figure}
   \plotone{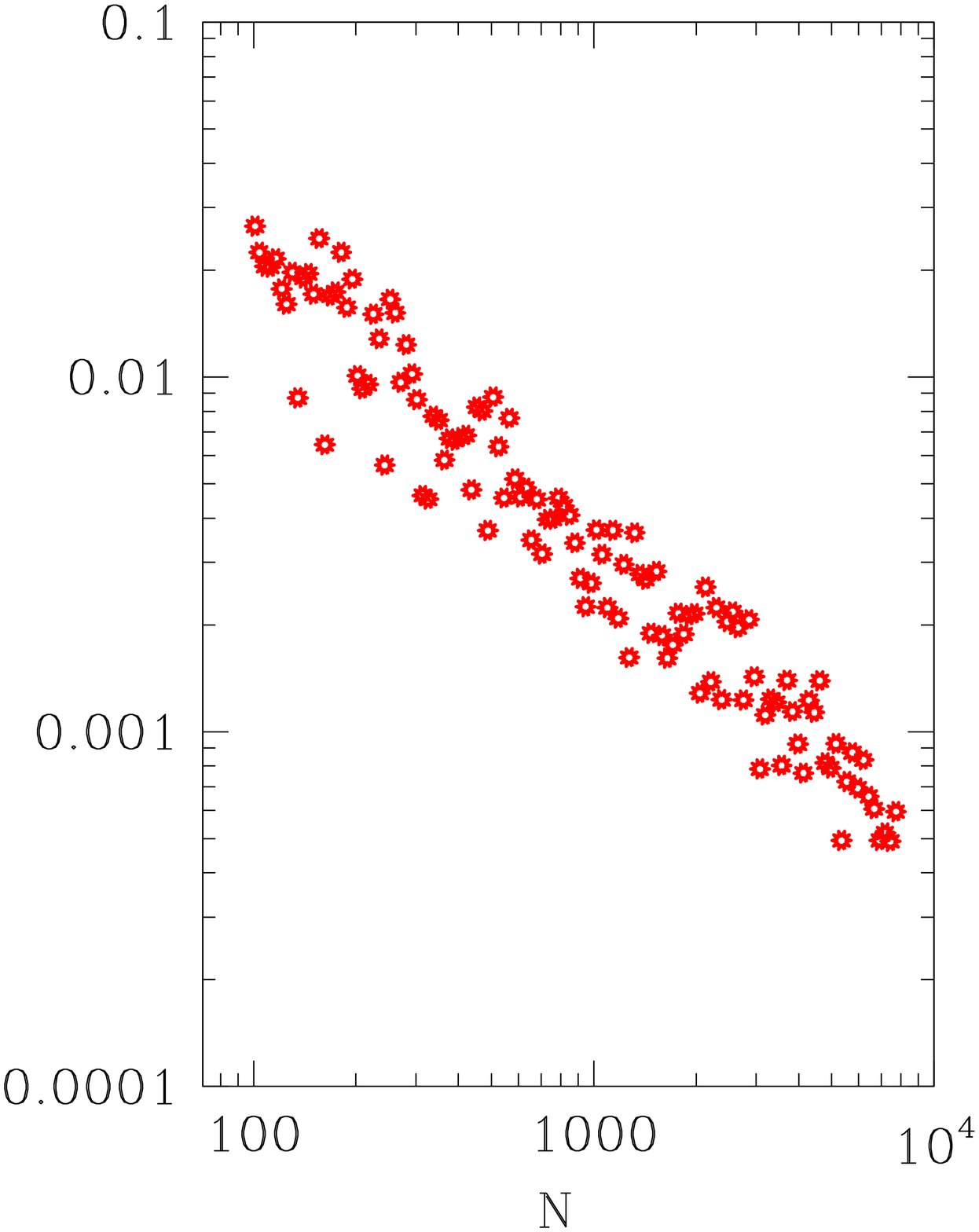}
\centerline{f11.ps}
\end{figure}
\clearpage

\begin{figure}
   \plotone{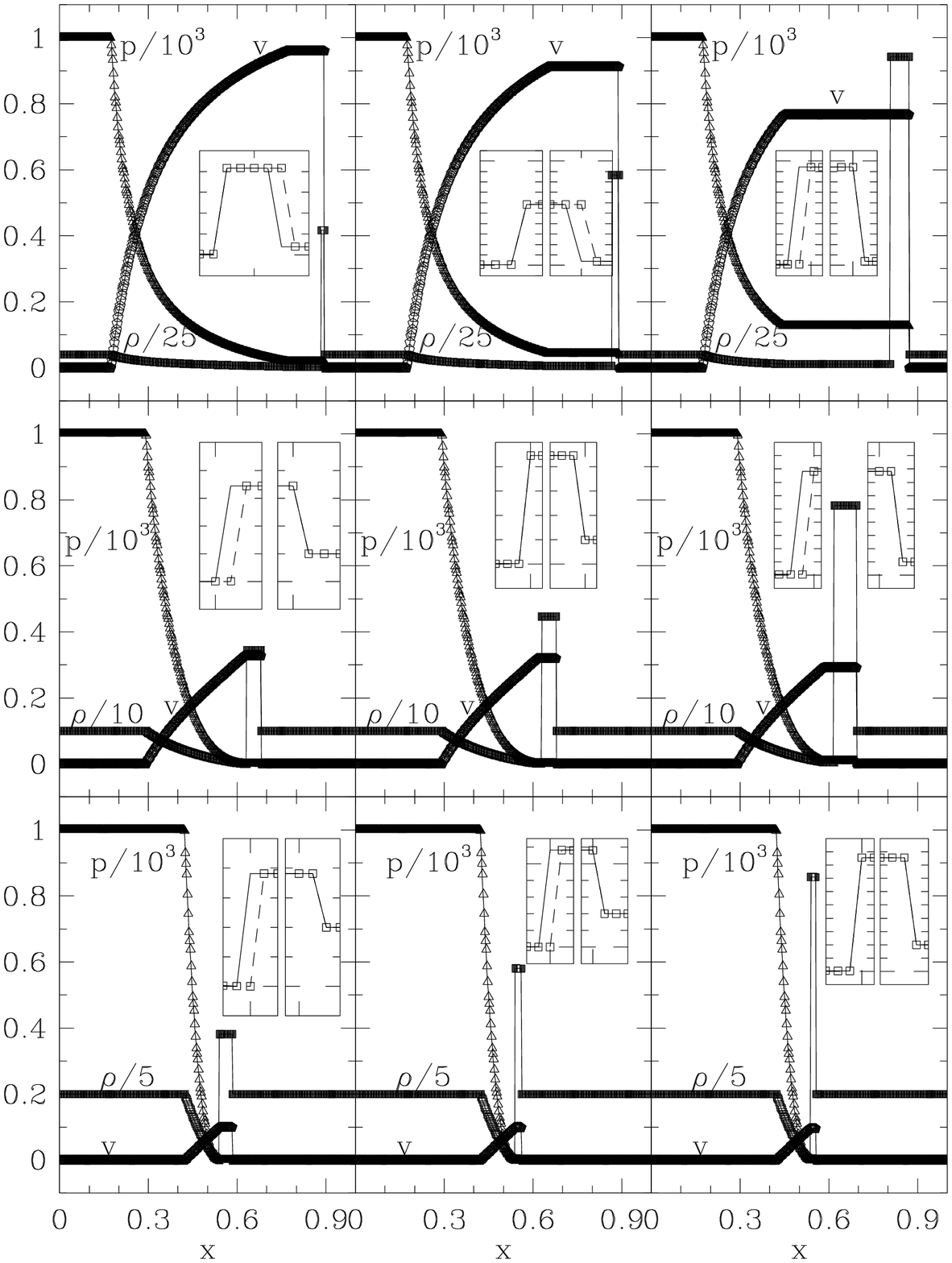}
\centerline{f12.ps}
\end{figure}
\clearpage

\begin{figure}
   \plotone{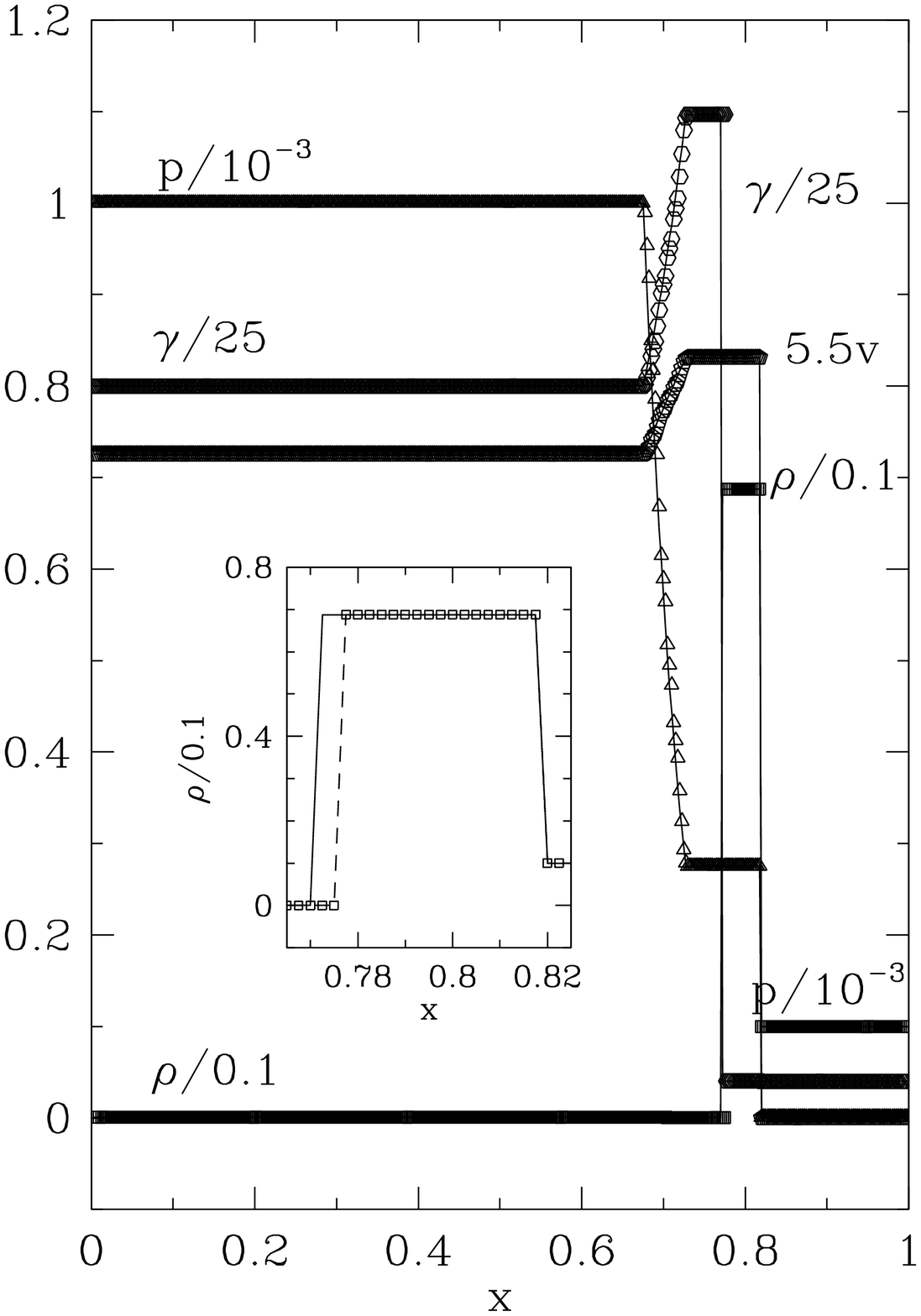}
\centerline{f13.ps}
\end{figure}
\clearpage

\begin{figure}
   \plotone{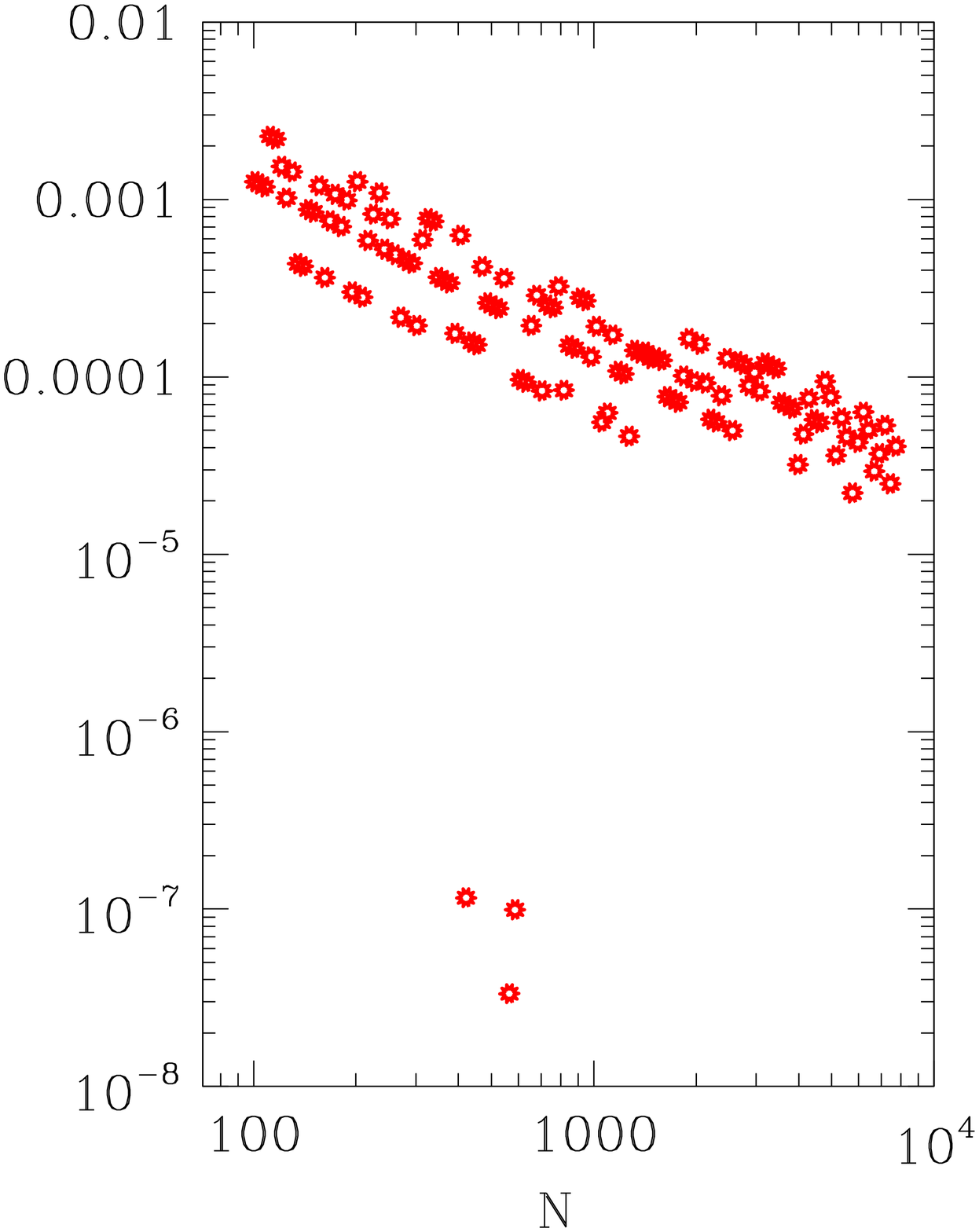}
\centerline{f14.ps}
\end{figure}
\clearpage

\begin{figure}
   \plotone{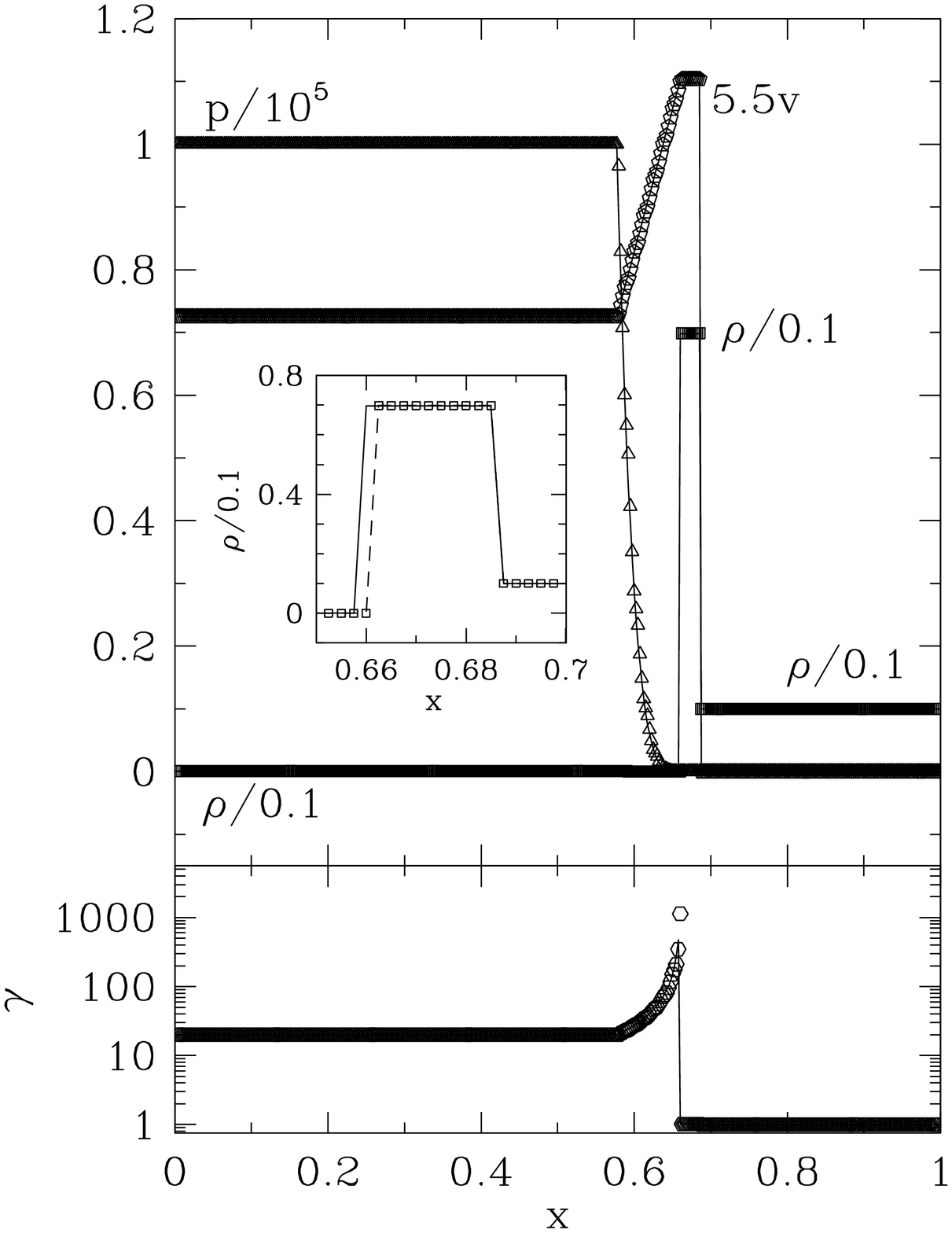}
\centerline{f15.ps}
\end{figure}
\clearpage

\begin{figure}
   \plotone{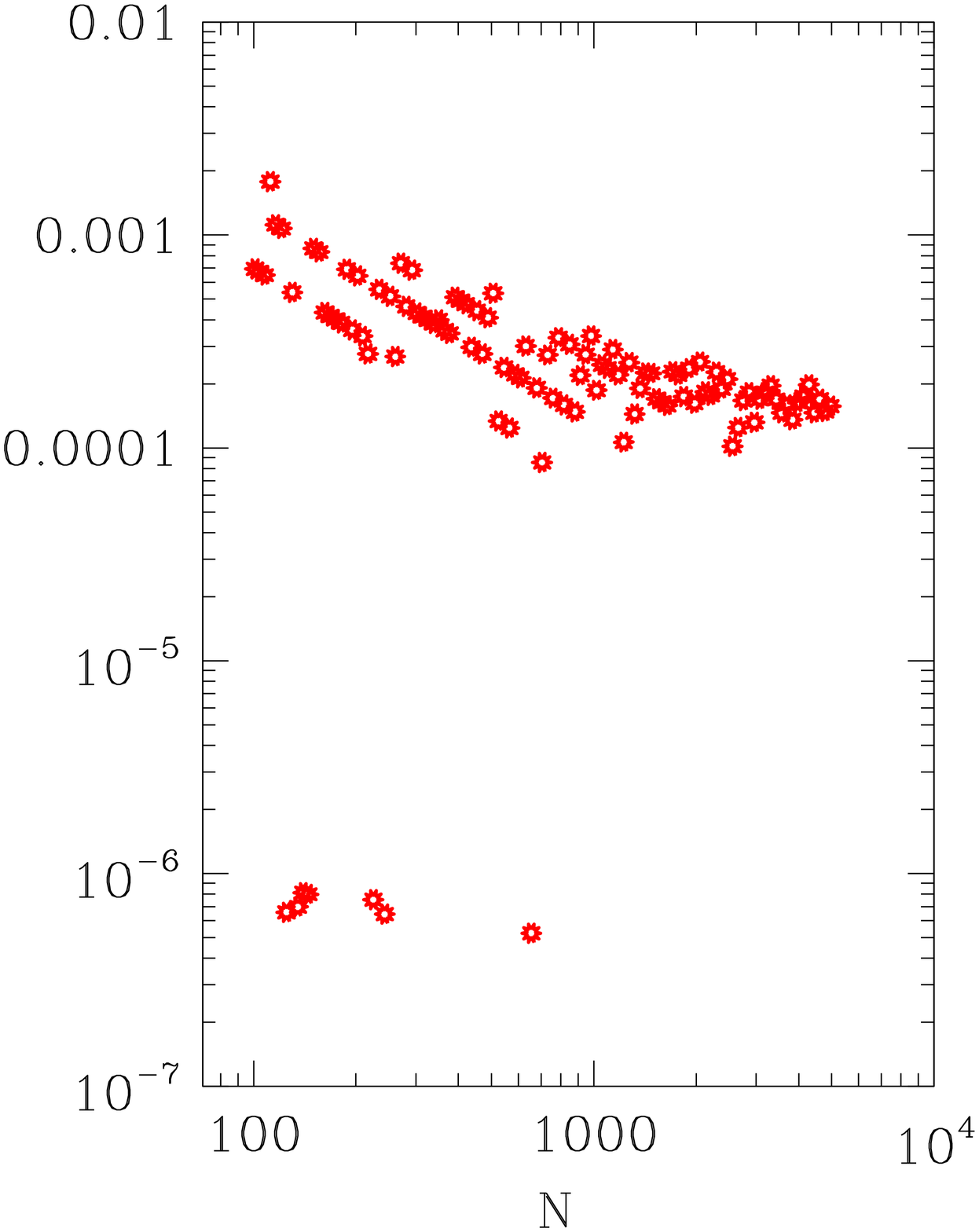}
\centerline{f16.ps}
\end{figure}
\clearpage

\begin{figure}
   \plotone{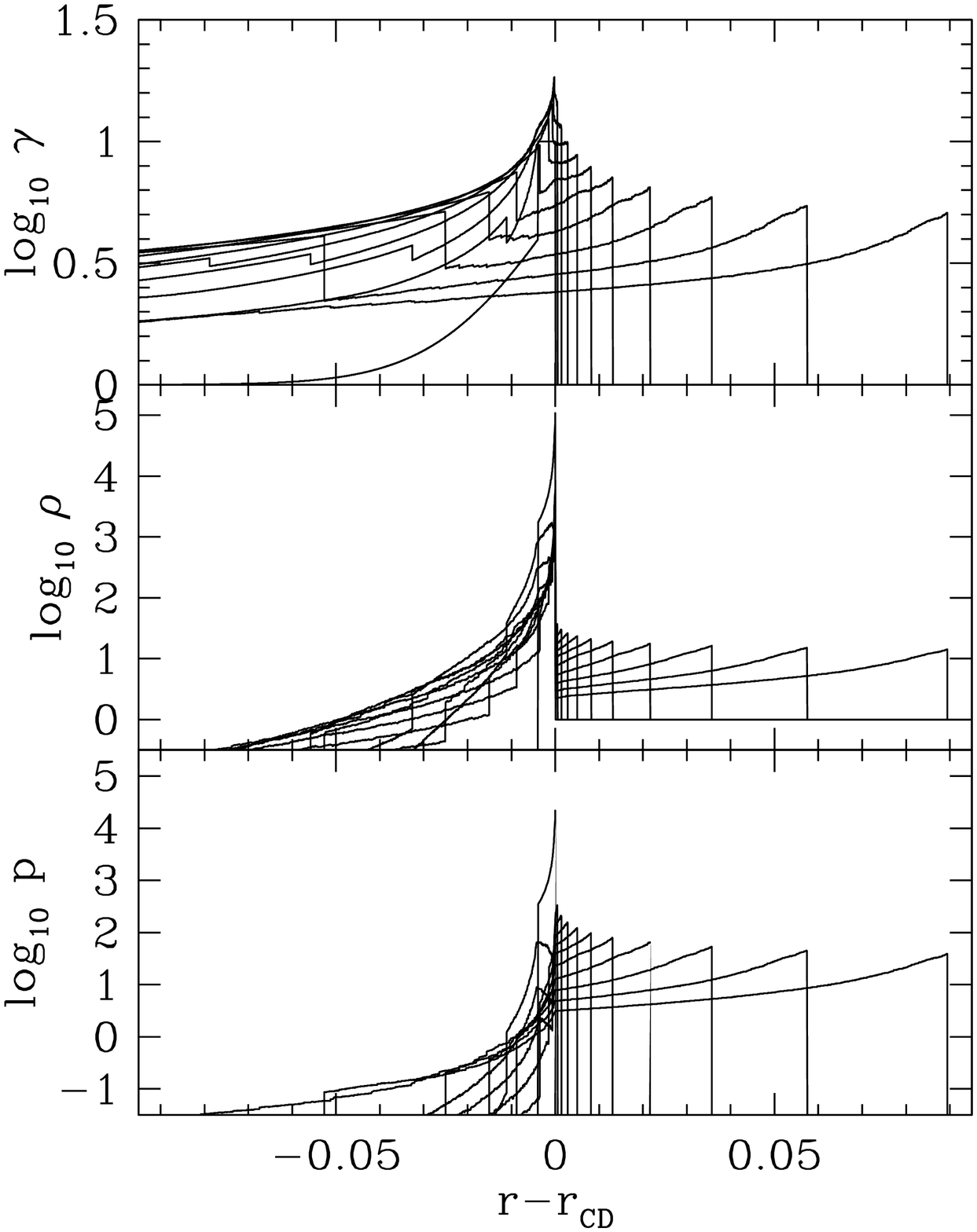}
\centerline{f17.ps}
\end{figure}
\clearpage

\begin{figure}
   \plotone{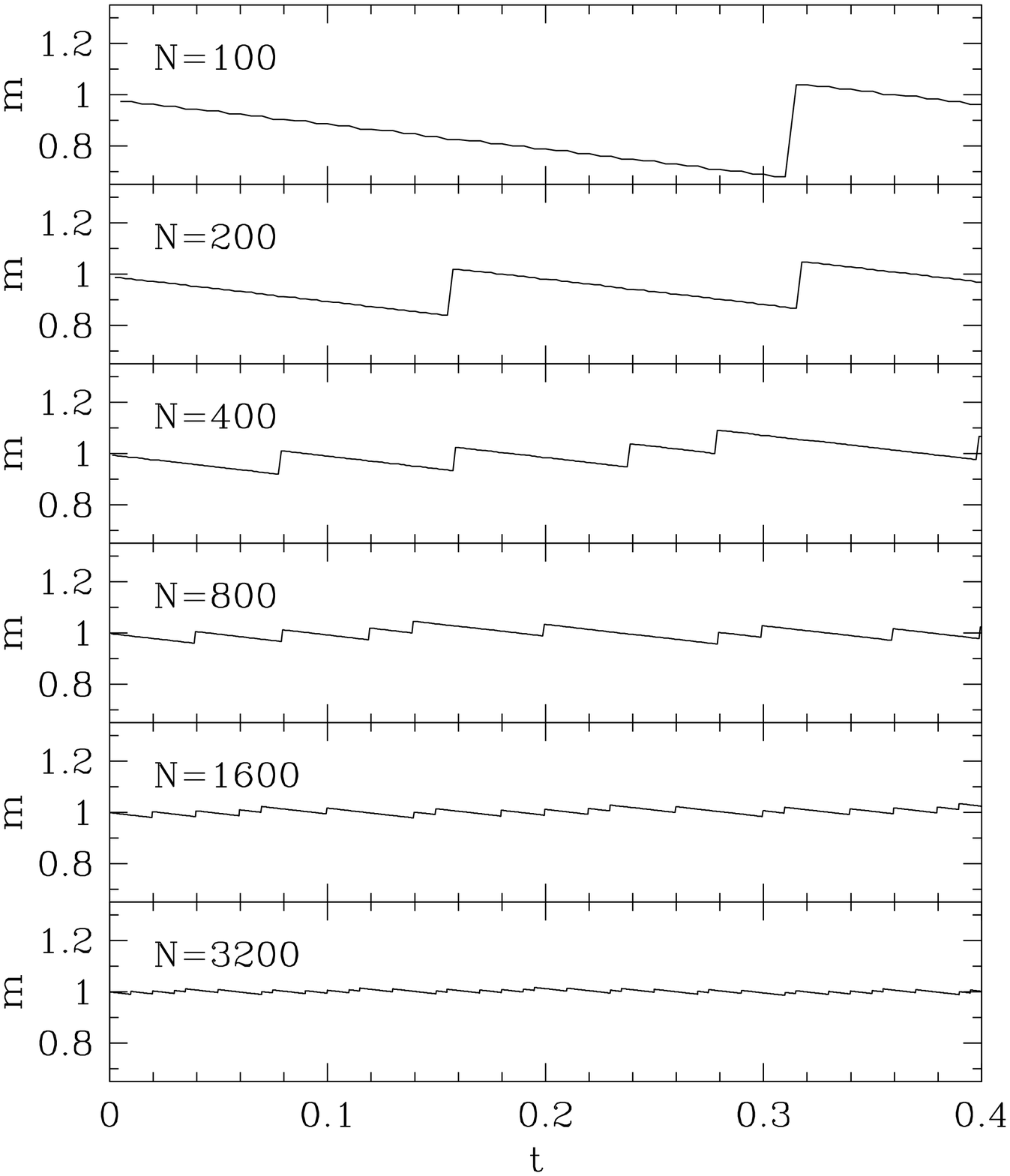}
\centerline{f18.ps}
\end{figure}
\clearpage

\end{document}